\documentclass[twocolumn,nolinenumbers]{aastex631}
\usepackage{enumitem}
\graphicspath{{./}{figures/}}
\usepackage[T1]{fontenc}
\usepackage{epsfig}
\usepackage{epstopdf}
\usepackage{natbib}
\usepackage{amssymb}
\usepackage{amsbsy}
\usepackage{natbib}
\usepackage{subfigure}
\usepackage[mathcal]{euscript}
\usepackage{float}
\usepackage{amsmath}
\usepackage{tabularx}
\usepackage{xspace}
\usepackage{enumitem}
\usepackage{xcolor}
\usepackage{comment}
\usepackage{hyperref}
\usepackage{ulem}

\usepackage{ulem,xspace}

\renewcommand{\Pr}{\ensuremath{\mathrm{Pr}}}
\newcommand{\Pm}{\ensuremath{\mathrm{Pm}}}

\newcommand{\Rm}{\mathrm{Rm}}

\newcommand{\uzrms}{\hat{u}_{z, \mathrm{rms}}}
\newcommand{\wrms}{\hat{u}_{z, \mathrm{rms}}}

\bibpunct{(}{)}{;}{a}{}{,}

\begin{document}

\title{Magnetized fingering convection in stars}

\author[0000-0003-4323-2082]{Adrian E.~Fraser}
\affiliation{Department of Applied Mathematics, University of Colorado, Boulder, CO 80309, USA}
\affiliation{Department of Astrophysical and Planetary Sciences, University of Colorado, Boulder, CO 80309, USA}
\affiliation{Laboratory for Atmospheric and Space Physics, University of Colorado, Boulder, CO 80303, USA}
\affiliation{Department of Applied Mathematics, Baskin School of Engineering, University of California, Santa Cruz, CA 95064, USA}

\author{ Sam A. Reifenstein}
\affiliation{Department of Applied Mathematics, Baskin School of Engineering, University of California, Santa Cruz, CA 95064, USA}

\author[0000-0002-6266-8941]{Pascale Garaud}
\affiliation{Department of Applied Mathematics, Baskin School of Engineering, University of California, Santa Cruz, CA 95064, USA}

\correspondingauthor{Adrian Fraser}
\email{adrian.fraser@colorado.edu}

\begin{abstract}
Fingering convection (also known as thermohaline convection) is a process that drives the vertical transport of chemical elements in regions of stellar radiative zones where the mean molecular weight increases with radius. 
Recently, \citet{Harrington_2019} used three-dimensional direct numerical simulations to show that a vertical magnetic field can dramatically enhance the rate of chemical mixing by fingering convection. Furthermore, they proposed a so-called "parasitic saturation" theory to model this process. 
Here, we test their model over a broad range of parameter space, using a suite of direct numerical simulations of magnetized fingering convection varying the magnetic Prandtl number, magnetic field strength, and composition gradient. 
We find that the rate of chemical mixing measured in the simulations is not always predicted accurately by their existing model, in particular when the magnetic diffusivity is large. We then present an extension of the \citet{Harrington_2019} model which resolves this issue. When applied to stellar parameters, it recovers the results of \citet{Harrington_2019} except in the limit where fingering convection becomes marginally stable, where the new model is preferred. We discuss the implications of our findings for stellar structure and evolution. 
\end{abstract}


\section{Introduction  } 
\label{sec:intro}
\setcounter{footnote}{0}
\subsection{Astrophysical motivation}

One of the original governing paradigms of stellar evolution models \citep{Eddington} is that stellar radiation zones are essentially laminar, because their strong stable density stratification quenches vertical turbulent motions. That paradigm has shifted over the last few decades, however, with the realization that there are a number of fluid instabilities that can develop and drive turbulence despite the stratification \citep[see the recent review by][and references therein]{garaud_journey_2021}. One such instability is the fingering instability (also called thermohaline instability), which can take place in the presence of what is often referred to as an "inverse" composition gradient, i.e., when the mean molecular weight of the star increases with radius. More precisely, a necessary condition for fingering convection \citep{Stern_1960,baines_gill_1969,ulrich_thermohaline_1972} is:
\begin{equation}
1 <  R_0 < \tau^{-1},
\end{equation}
where the diffusivity ratio $\tau = \kappa_C /\kappa_T$ is the ratio of the microscopic diffusivity $\kappa_C$ of the chemical species which most contributes to the inverse composition gradient, to the thermal diffusivity $\kappa_T$, and where the density ratio
\begin{equation}
R_0 = \frac{ | N_T^2 | }{|N_C^2|}
\end{equation}
is the ratio of the stable temperature stratification (measured by $N_T^2>0$, the square of the Brunt-V\"ais\"al\"a frequency associated with the temperature stratification only) to the unstable composition stratification (measured by $N_C^2<0$, the corresponding quantity for composition). 
In stellar interiors, the thermal diffusivity is significantly greater than the compositional diffusivity so $\tau^{-1} \gg 1$, and thus even very slight inverse composition gradients may drive fingering convection. 

A wide variety of stars are believed to undergo fingering convection, with consequences that are possibly observable, or may otherwise impact their evolution or the interpretation of observations. In low-mass red giant branch (RGB) stars, for example, extra mixing is required to explain abundance observations near the luminosity bump \citep[e.g.,][]{Gratton, shetrone_constraining_2019} and is largely thought to be due to fingering convection \citep[first noted by][see also \citealt{Baltimore_paper} and references therein]{CharbonnelZahn2007}. 
The surface abundances of metal-rich white dwarfs (WDs) are often thought to indicate active accretion \citep{jura_tidally_2003,farihi_infrared_2009,Koester}. This creates an unstable composition gradient near the surface  that can drive fingering convection. The enhanced turbulent mixing rate must be taken into account in order to infer the accretion rate from observations \citep{Deal_WD,bauer_increases_2018,bauer_polluted_2019,Wachlin_WDs}. 
\textcolor{black}{Also in WDs, 
a destabilizing composition gradient is thought to form when cooling drives crystalization in the core \citep{stevenson_eutectic_1980,Mochkovitch}. The potential Rayleigh-Taylor instability of this gradient has been put forth as a possible source of dynamo activity to explain observed WD magnetic fields \citep{isern_common_2017,ginzburg_slow_2022}; recent work by \citet{fuentes_heat_2023} and \citet{Montgomery}, however, has demonstrated that this configuration is likely fingering-unstable, with important consequences to the possibility of a crystallization-driven dynamo.}
Fingering convection is also believed to occur in the envelopes of accretor stars in massive stellar binaries, and to significantly affect their interior composition profiles \citep{Renzo}. 
Similarly, fingering convection is thought to be an important process in the formation of carbon-enhanced metal-poor stars by accretion of carbon-rich material from a massive, evolved companion \citep{stancliffe_carbon-enhanced_2007}.

Quantifying the precise impact of fingering convection on stellar evolution requires a model for mixing by the turbulence it creates. Unfortunately, despite the enormous progress made in supercomputing in the past 30 years, it is still not possible to perform three-dimensional direct numerical simulations (DNS) of fingering convection at the parameters appropriate for stellar interiors. Instead, one must rely on simple ad-hoc models, tested (whenever possible) using DNS at parameters that are achievable numerically. We now describe recent progress on that front.
 
\subsection{From traditional to modern models of fingering convection}

Early prescriptions for turbulent mixing by fingering convection based on dimensional analysis date back as far as \citet{ulrich_thermohaline_1972} and \citet{kippenhahn_thermohaline}. Both studies argued that the turbulent compositional flux can be modeled as a diffusive process, with a diffusion coefficient $D_\mathrm{mix}$ that can be written as
\begin{equation}\label{eq:kippenhahn}
    D_\mathrm{mix} = C_t \kappa_T / R_0,
\end{equation}
where $C_t$ is a free parameter. 
Plausible values for $C_t$ put forth in these original calculations vary by orders of magnitude, from $C_t = O(1000)$ \citep{ulrich_thermohaline_1972} to $C_t = O(10)$ \citep{kippenhahn_thermohaline}. 
\citet{CharbonnelZahn2007} found $C_t = O(1000)$ to be necessary in order for their 1D stellar evolution models of RGB stars to achieve mixing levels consistent with the observations of \citet{Gratton}. This was later confirmed by a similar study of \citet{Denissenkov} \citep[for a review of efforts to match observations using this model, see][]{Salaris_review}.

Since 2010, progress in high-performance computing has enabled us to estimate compositional fluxes in fingering convection using three-dimensional DNS in numerically tractable parameter regimes and test these models \citep{garaud_DDC_review}. In particular, \citet{Traxler2011a} and \citet{Brown_2013} found that the fluxes predicted by \citet{ulrich_thermohaline_1972} overestimate the results of their hydrodynamic DNS by very large amounts. 
As a result, fingering convection alone does not seem to be a strong enough source of mixing to explain observations of RGB star abundances according to the results of \citet{CharbonnelZahn2007}. Furthermore, the functional dependence of the turbulent diffusivity on $R_0$ and $\tau$ predicted by Eq.~\eqref{eq:kippenhahn} is not consistent with DNS results, prompting a search for new, better models. 

\citet{Brown_2013}, following similar work in the oceanographic context by \citet{RadkoSmith2012} \citep[which itself built upon the pioneering work by][]{holyer_1984}, developed a "parasitic saturation" model that agrees much more favorably with their simulations. Generally speaking, parasitic models (described in greater detail in Sec.~\ref{summary:subsec:parasites} below) assume that a given primary instability saturates nonlinearly because of the development of secondary "parasitic instabilities" that feed on their energy \citep[for applications in a variety of contexts beyond fingering convection, see, e.g.,][]{Goodman_Xu,Pessah_Goodman,Pessah,latter_mri_2009,longaretti_mri-driven_2010,barker2019,barker2020}. In the case of fingering convection, \citet{RadkoSmith2012} and \citet{Brown_2013} demonstrated that shear instabilities can cause the saturation of the fingering instability when their respective growth rates are similar. Applying the theory yields estimates for the vertical velocity of the fingers at saturation, which can then be used to estimate the vertical heat and compositional transport by fingering convection.

The parasitic model of \citet{Brown_2013} correctly predicts the turbulent compositional fluxes measured in hydrodynamic DNS of fingering convection for a wide range of input parameters \citep[see, e.g., Figure 2 of][]{garaud_DDC_review}, suggesting that one could confidently use it at stellar parameters as well. Doing so, however, continues to predict compositional fluxes that are much smaller than those required to explain abundance observations in RGB stars \citep{CharbonnelZahn2007}, and potentially challenges the notion that this instability is the answer to this particular observational conundrum. And yet, the fact that extra mixing is needed precisely at the time when an inverse composition gradient forms in the star \citep[see counter-arguments by][however]{tayar_joyce_2022}, and that the extra mixing required increases with the magnitude of the composition gradient \citep{Baltimore_paper}, as one would expect in fingering convection, makes it very difficult to abandon the idea of fingering convection altogether. 

\subsection{Magnetized fingering convection}

\citet{Harrington_2019} (HG19 hereafter) recently demonstrated using DNS 
that a uniform, vertical magnetic field (that is aligned with the direction of gravity) can increase transport in fingering convection by suppressing the parasitic shear instabilities and letting finger velocities grow to larger amplitudes before saturation. 
They also extended the \citet{Brown_2013} model by accounting for the magnetic field in the calculation of the growth rates of the parasitic shear instabilities. They found that this new magnetohydrodynamic (MHD) parasitic model correctly accounts for the increase in compositional flux with magnetic field strength observed in their DNS. 
The HG19 model predicts that magnetic field strengths as low as $O(100 \text{G})$ could increase the efficiency of fingering convection in RGB stars by two orders of magnitude, thus potentially resolving the RGB stars abundance conundrum discussed above. 

Despite these promising results, it is important to note that the HG19 work was very preliminary and only explored a very limited range of numerically accessible input parameters: they fixed $R_0 = 1.45$ and $\Pr = \tau = 0.1$, where $\Pr$ is the Prandtl number (the ratio of the kinematic viscosity to the thermal diffusivity), and used a magnetic Prandtl number $\Pm = 1$, where $\Pm$ is the ratio of kinematic viscosity to magnetic diffusivity. These choices are, in hindsight, somewhat problematic. Indeed, the radiative zones of many stars usually have $\Pm < 1$. In addition, fingering regions likely span a wide range of density ratios $R_0$ \citep[see, e.g.,][]{Baltimore_paper}. Thus, the HG19 model merits comparison with DNS over a much wider range of parameters than what has been covered so far. 

\subsection{Structure of the paper}
In this paper we extend the work of HG19 in two ways. First, we perform DNS over a broader range of density ratios $R_0$ and at $\Pm < 1$ (using the same uniform, vertical background magnetic field -- we discuss some caveats pertaining to this choice in Sec.~\ref{sec:conclusions}). We find that there are regimes where the HG19 model fails because viscosity and resistivity, which were neglected by HG19, begin to dominate the dynamics of the parasitic instability. 
Second, we extend the HG19 model in an attempt to account for these effects, and compare the results to our DNS. We first add viscosity and resistivity in the computation of the parasitic instability properties, but find that this is unable to predict the fluxes obtained in the DNS. Then, following \citet{RadkoSmith2012}, we also add the temperature and composition fields to the parasitic instability model, and find that it now matches the data with high fidelity.

The paper is outlined as follows: we introduce our governing equations, non-dimensionalization, and review the basic properties of the fingering instability in Sec.~\ref{sec:DNS_setup}, followed by a summary of previous work in Sec.~\ref{sec:summary}. In Sec.~\ref{sec:DNS_results}, we present the results of our DNS and compare them against the HG19 model. We describe our extension to the HG19 model and compare it to our DNS in Sec.~\ref{sec:extending_HG19}. Our conclusions are presented in Sec.~\ref{sec:conclusions}.

\section{Problem setup }
\label{sec:DNS_setup}
\subsection{Governing equations} \label{setup:subsec:PDEs}

We study fingering convection in MHD using the same model as HG19, with the key features outlined here for clarity. 
We assume that the typical length scales of turbulent fluctuations are much shorter than the temperature, density and pressure scale heights, and that typical velocity fluctuations are much smaller than the local sound speed. These assumptions allow us to use the Boussinesq approximation for compressible gases \citep{spiegel_boussinesq_1960}. Similarly, we assume that the relevant length scales are much smaller than the radius of the star to justify the use of a Cartesian grid $(x, y, z)$, with $z$ taken as the vertical (i.e., radial) direction. 
The governing equations are then
\begin{equation}
    \nabla \cdot \mathbf{u} = 0,
\end{equation}
\begin{multline}
    \rho_m \left( \frac{\partial \mathbf{u}}{\partial t} + \mathbf{u} \cdot \nabla \mathbf{u} \right) = - \nabla p + \rho_m \nu \nabla^2 \mathbf{u} \\ + \frac{1}{\mu_0} \left( \nabla \times \mathbf{B} \right) \times \mathbf{B} + \rho_m \left(-\alpha T + \beta C\right) \mathbf{g},
\end{multline}
\begin{equation}
    \frac{\partial T}{\partial t} + \mathbf{u} \cdot \nabla T + u_z \left( \frac{dT_0}{dz} - \frac{dT_\mathrm{ad}}{dz} \right) = \kappa_T \nabla^2 T,
\end{equation}
\begin{equation}
    \frac{\partial C}{\partial t} + \mathbf{u} \cdot \nabla C + u_z \frac{dC_0}{dz} = \kappa_C \nabla^2 C,
\end{equation}
\begin{equation}
    \frac{\partial \mathbf{B}}{\partial t} = \nabla \times \left( \mathbf{u} \times \mathbf{B} \right) + \eta \nabla^2 \mathbf{B},
\end{equation}
and
\begin{equation}
    \nabla \cdot \mathbf{B} = 0.
\end{equation}
In these equations $\mathbf{u} = (u_x, u_y, u_z)$ is the velocity field, $\mathbf{B} = (B_x, B_y, B_z)$ is the magnetic field, $p$ is pressure perturbation away from hydrostatic equilibrium (which is assumed for the background stratification), and $T$ and $C$ are the temperature and composition perturbations away from assumed linear background profiles with gradients $dT_0/dz$ and $dC_0/dz$, respectively. All other quantities are model constants (consistent with the use of the Boussinesq approximation). The mean density of the domain is $\rho_m$, and $\alpha$ and $\beta$ are the coefficients of thermal expansion and compositional contraction, respectively. 
The adiabatic temperature gradient is $dT_\mathrm{ad} / dz = -g / c_p$, where $g = |\mathbf{g}|$ is the local gravity and $c_p$ is the specific heat at constant pressure. 
In stellar radiative zones, $dT_\mathrm{ad}/dz < dT_0/dz<0$.
The kinematic viscosity, thermal diffusivity, compositional diffusivity, and magnetic diffusivity are given by $\nu$, $\kappa_T$, $\kappa_C$, and $\eta$, respectively, and are also taken to be constant. 
Finally, $\mu_0$ is the permeability of free space. 
Note that magnetic buoyancy is not included in this model. 

The dynamical fields of interest, $\mathbf{u}$, $\mathbf{B}$, $T$, and $C$, are assumed to satisfy periodic boundary conditions along each axis, a convenient property for direct numerical simulations because it enables the use of Fourier series expansions in pseudospectral codes, and avoids unphysical boundary layers that inevitably occur near impermeable boundaries. 

We then non-dimensionalize the equations as in HG19. 
Length scales are measured in units of the characteristic width of fingers \citep{Stern_1960},
\begin{equation} \label{eq:x-units}
    d = \left( \frac{\kappa_T \nu}{\alpha g \left( \frac{dT_0}{dz} - \frac{dT_\mathrm{ad}}{dz} \right)} \right)^{1/4} = \left( \frac{\kappa_T \nu}{N_T^2} \right)^{1/4},
\end{equation}
where $N_T$ is the local Brunt–V\"{a}is\"{a}l\"{a} frequency defined from the temperature stratification. 
The other units for time, velocity, temperature,  composition, and magnetic field are, respectively, 
\begin{equation} \label{eq:t-units}
    [t] = \frac{d^2}{\kappa_T}, [v] = \frac{\kappa_T}{d},
\end{equation}
\begin{equation}
    [T] = d \left( \frac{dT_0}{dz} - \frac{dT_\mathrm{ad}}{dz} \right), 
    [C] = \frac{\alpha}{\beta} [T],
\end{equation}
and
\begin{equation} \label{eq:b-units}
    [B] = B_0,
\end{equation}
where $B_0$ is the amplitude of an assumed mean vertical field (see Sec.~\ref{sec:intro} and below). 

The non-dimensional governing equations are (cf. HG19)
\begin{multline} \label{eq:dimless-mom}
    \frac{\partial \hat{\mathbf{u}}}{\partial t} + \hat{\mathbf{u}} \cdot \nabla \hat{\mathbf{u}} = - \nabla \hat{p} + \Pr \nabla^2 \hat{\mathbf{u}} \\ + H_B \left( \nabla \times \hat{\mathbf{B}} \right) \times \hat{\mathbf{B}} + \Pr \left( \hat{T} - \hat{C} \right) \hat{\mathbf{e}}_z,
\end{multline}
\begin{equation}
    \frac{\partial \hat{T}}{\partial t} + \hat{\mathbf{u}} \cdot \nabla \hat{T} + \hat{u}_z = \nabla^2 \hat{T},
\end{equation}
\begin{equation}
    \frac{\partial \hat{C}}{\partial t} + \hat{\mathbf{u}} \cdot \nabla \hat{C} + \frac{\hat{u}_z}{R_0} = \tau \nabla^2 \hat{C},
\end{equation}
\begin{equation} \label{eq:dimless-induction}
    \frac{\partial \hat{\mathbf{B}}}{\partial t} = \nabla \times \left( \hat{\mathbf{u}} \times \hat{\mathbf{B}} \right) + D_B \nabla^2 \hat{\mathbf{B}},
\end{equation}
\begin{equation}
    \nabla \cdot \hat{\mathbf{u}} = 0,
\end{equation}
and
\begin{equation} \label{eq:dimless-divB}
    \nabla \cdot \hat{\mathbf{B}} = 0,
\end{equation}
where hats over dynamical fields indicate non-dimensional quantities, and $\hat{\mathbf{e}}_z$ is the unit vector in the $z$ direction. 
From here on, unless otherwise stated, space and time coordinates are implicitly assumed to be non-dimensionalized in this way. 
The system is now specified by five dimensionless parameters: 
\begin{align}
    \Pr &= \frac{\nu}{\kappa_T}, \quad \tau = \frac{\kappa_C}{\kappa_T}, \quad D_B = \frac{\eta}{\kappa_T}, \nonumber \\
    R_0 &= \frac{\alpha \left( \frac{dT_0}{dz} - \frac{dT_\mathrm{ad}}{dz} \right)}{\beta \frac{dC_0}{dz}}, \quad H_B = \frac{B_0^2 d^2}{\rho_m \mu_0 \kappa_T^2},
\end{align}
where $\Pr$ is the Prandtl number, $\tau$ and $D_B$ are the compositional and resistive diffusivity ratios, respectively, $R_0$ is the density ratio, and $H_B$ is the coefficient of the Lorentz force, which can be understood as the squared inverse of the Alfv\'{e}n Mach number based on the unit magnetic field and the unit flow speed. 
In stars, the parameters $\Pr$, $\tau$, and $D_B$ are all extremely small \citep[see, e.g.,][]{Garaud2015}. 
The density ratio characterizes the stabilizing influence of the stable temperature gradient relative to the destabilizing composition gradient, and, in radiation zones, can take any value from 1 to infinity depending on local stellar conditions. The threshold $R_0 = 1$ is equivalent to the Ledoux criterion for convection, and $R_0>1$ applies to regions that are stable to convection but potentially unstable to fingering convection, as described below in Sec.~\ref{setup:subsec:linear}.
Throughout this paper, we will often refer to the magnetic Prandtl number, defined as
\begin{equation}
    \Pm = \frac{\nu}{\eta} = \frac{\Pr}{D_B}.
\end{equation}
This ratio is typically $O(10^{-2})-O(1)$ in the radiative zones of most stars. 

\subsection{Linear stability analysis} \label{setup:subsec:linear}

A thorough discussion of the linear stability properties of non-magnetic fingering convection at low $\Pr$ (the regime relevant to stellar interiors) can be found in \citet{Brown_2013}, with the magnetized case first explored in \citet{Harrington_thesis} and  HG19. 
The most important result is that for a given diffusivity ratio $\tau$, fingering convection takes place when $1 < R_0 < 1/\tau$, with the system becoming less unstable as $R_0$ increases \citep{Stern_1960,baines_gill_1969}. 

More specifically, eigenmode solutions to the linearized system for small-amplitude perturbations take the form
\begin{equation} \label{eq:normal-mode}
    \hat{q} = \hat{q}_0 \exp[\hat{\lambda}t + i(\hat{\mathbf{k}} \cdot \mathbf{x})],
\end{equation}
where $q$ represents each of the dynamical fields, $\hat{\mathbf{k}} = (\hat{k}_x, \hat{k}_y, \hat{k}_z)$ is the mode's wavenumber, and $\hat{\lambda}$ is its growth rate. The fastest-growing modes are so-called ``elevator modes", with a velocity that is in the vertical direction only, and where all perturbations are invariant in $z$ (so $\hat{k}_z = 0$). 
\citet{Harrington_thesis} and HG19 showed that the same is true for magnetized fingering convection with a uniform background magnetic field regardless of its orientation. 
When the magnetic field is vertical, which is the case considered here, the field lines are parallel to the fluid flow within each elevator, and therefore do not interact with the latter. 
As a result, the properties of the fastest-growing modes (growth rate and wavenumber) are identical in MHD and in the hydrodynamic limit. 
The magnetic field, however, stabilizes fingering modes with $\hat k_z \ne 0$ and also plays a crucial role in the nonlinear saturation of the elevators/fingers by stabilizing them against secondary shear instabilities (see HG19). This in turn controls the typical velocity of the fluid within the fingers after saturation, and sets the transport rate of heat and chemical species by the fingering convection.

\subsection{Turbulent diffusion by fingering convection} 

The rate at which a turbulent fluid transports temperature and composition is quantified by the thermal and compositional fluxes, defined (dimensionally) as
\begin{equation} \label{eq:FT_def}
    F_T = \langle {u}_z {T} \rangle,
\end{equation}
and
\begin{equation} \label{eq:FC_def}
    F_C = \langle {u}_z {C} \rangle,
\end{equation}
respectively, where $\langle . \rangle$ denotes a volume average as well as a time average taken once the turbulence has reached a statistically stationary state (see Sec.~\ref{DNS:subsec:methods}).
Note that both fluxes are negative in fingering convection \citep{radko_book}. In stellar evolution, it is customary to assume that the turbulent compositional flux can be modeled as a turbulent diffusion process, with a turbulent diffusivity $D_C$ defined (dimensionally) such that
\begin{equation}
\label{eq:DCdef}
     F_C = - D_C \frac{dC_0}{dz} \leftrightarrow D_C = - \frac{\langle {u}_z {C} \rangle}{\frac{dC_0}{dz}} .
\end{equation}
In the units of Section \ref{setup:subsec:PDEs}, this becomes
\begin{equation}
    \hat D_C  =\frac{D_C}{\kappa_T} = - R_0 \hat F_C  = - R_0 \langle \hat{u}_z \hat{C} \rangle . 
\end{equation}
In other words, modeling the turbulent diffusion coefficient $D_C$ for magnetized fingering convection in stars requires estimating the nondimensional flux $\hat F_C$ as a function of the local properties of the star, which control the non-dimensional input parameters of the model, namely $\Pr$, $\tau$, $R_0$, $H_B$, and $D_B$. In what follows, we first summarize recent preliminary results in that direction obtained by HG19, who used a combination of numerical simulations and semi-analytical modelling to predict $\hat F_C$ for a narrow range of parameter space. We then expand their work to a much wider range of parameter space.

\section{Summary of previous work}
\label{sec:summary}
\subsection{Previous numerical experiments} 

The first DNS of fingering convection in the presence of a magnetic field, solving equations \eqref{eq:dimless-mom}-\eqref{eq:dimless-divB}, were presented by \citet{Harrington_thesis} and HG19. 
The majority of their simulations assumed a uniform  vertical background magnetic field, as we do here, with an amplitude selected so that $H_B$ spans the interval $[10^{-2},10^2]$. 
In that preliminary work, all the other parameters were held constant, with the diffusivity ratios fixed to be $\Pr = \tau = 0.1$ and $\Pm = 1$, and the density ratio set to $R_0 = 1.45$. 
As discussed in Sec.~\ref{sec:intro}, HG19 found that increasing the amplitude of the background magnetic field can significantly enhance the magnitude of the turbulent vertical fluxes of both heat and composition, a result they explained by noting that the field stabilizes the secondary shear instability that normally saturates the primary fingering instability, letting the fingers grow to larger amplitude before saturation. In order to quantify this idea, they proposed a new parasitic saturation model for fingering convection in MHD that is built upon the one successfully presented by \citet{Brown_2013} in the hydrodynamic limit. 

\subsection{Parasitic saturation models} \label{summary:subsec:parasites}

As described in Sec.~\ref{sec:intro}, parasitic saturation models assume that the primary fingering instability (which is dominated by the fastest-growing elevator modes) saturates when secondary shear instabilities developing between the upflowing and downflowing elevators take over. If we let $\hat w_f$ and $\hat C_f$ be the amplitude of the vertical velocity and composition perturbation in the fastest-growing elevator mode at saturation, then the corresponding turbulent compositional flux can reasonably be assumed to be proportional to the product $\hat w_f\hat C_f$ \citep{RadkoSmith2012,Brown_2013}, i.e.,
\begin{equation} \label{eq:ingredient1}
    {\hat F}_C = C_1 \hat{w}_f \hat{C}_f = - C_1 \frac{\hat{w}_f^2}{R_0(\hat{\lambda}_f + \tau \hat{l}_f^2)},
\end{equation}
where $C_1$ is a model parameter that is assumed to be a universal constant, $\hat{\lambda}_f$ and $\hat{l}_f$ are the nondimensional growth rate and horizontal wavenumber of the fastest-growing elevator mode, and where in the second equality, $\hat{C}_f$ has been expressed in terms of $\hat{w}_f$ by using the linearized equations described in Sec.~\ref{setup:subsec:linear} \citep[see][Sec.~4.2 for details]{Brown_2013}. Note that $\hat F_C < 0$, as required; in addition, note that this expression is only valid in the fingering range, i.e. for $1 < R_0 < \tau^{-1}$. 

The parasitic saturation condition is expressed mathematically as a rough balance between the growth rate of the fingering elevator mode $\hat{\lambda}_f$ and the growth rate of the fastest-growing parasitic mode $\hat{\sigma}$, namely
\begin{equation} \label{eq:ingredient2}
    \hat{\lambda}_f = C_2 \hat{\sigma},
\end{equation}
where $C_2$ is another model parameter assumed to be constant. Because shear instabilities almost always grow faster when the shear is stronger, $\hat{\sigma}$ is (usually) a monotonically increasing function of $\hat{w}_f$. 
Eq.~\eqref{eq:ingredient2} can then be inverted  to obtain $\hat{w}_f$ as a function of $\hat \lambda_f$, which then yields $\hat F_C$ as a function of known quantities only via Eq.~\eqref{eq:ingredient1}. 

In the hydrodynamic case, \citet{Brown_2013} ignored the effect of viscosity on the development of the shear instability, in which case the computation of $\hat \sigma$ can be done semi-analytically to yield $\hat \sigma = K \hat{w}_f \hat{l}_f$, where $K \simeq 0.26$ is a known constant that is obtained numerically by solving the linear stability problem. Combining this result with Eq. (\ref{eq:ingredient2}) yields
\begin{equation} \label{eq:brown_wf}
\hat w_f = (C_2K)^{-1} \frac{\hat \lambda_f}{\hat l_f},
\end{equation}
and provides an estimate of the vertical velocity of the elevator modes at saturation. \citet{sengupta_2018} compared this expression with the measured $\wrms$ in DNS to estimate the prefactor as $(C_2 K)^{-1} \simeq 2\pi$, and hence $C_2 \simeq 0.6$. Finally, substituting this into Eq.~\eqref{eq:ingredient1} yields
\begin{equation}
    \hat F_C = - \left( \frac{C_1}{C_2^2 K^2} \right) \frac{\hat{\lambda}_f^2}{\hat{l}_f^2R_0(\hat{\lambda}_f + \tau \hat{l}_f^2)},
\end{equation}
which was fitted by \citet{Brown_2013} against the available hydrodynamic data for $\hat F_C$ to estimate $C_1(C_2 K)^{-2} \simeq 49$, and thus $C_1 \simeq 1.24$. This expression, as demonstrated by \citet{Brown_2013} and \citet{garaud_DDC_review}, provides a very good estimate of the compositional flux $\hat F_C$ over most of parameter space.

In the magnetic case, the shear instability growth rate $\hat \sigma$ also depends on the magnetic field strength, via $H_B$, and (possibly) on the magnetic diffusivity, via $D_B$. Following \citet{Brown_2013}, HG19 chose to consider the so-called "ideal MHD" limit where now both viscosity and resistivity are neglected in the calculation of $\hat{\sigma}$. They showed that, in this limit, the growth rate of the dominant parasitic mode, normalized to the elevator mode amplitude and wavenumber, namely
\begin{equation} \label{eq:sigma_star}
\sigma^* \equiv \frac{\hat{\sigma}}{(\hat{w}_f \hat{l}_f)},
\end{equation}
is purely a function of 
\begin{equation} \label{eq:HB_star}
H_B^* \equiv \frac{H_B}{\hat{w}_f^2},
\end{equation}
which is the squared ratio of the Alfv\'{e}n velocity to the saturated elevator mode velocity. Moreover, they found that this function is approximated well by the expression
\begin{equation} \label{eq:HG19_sigma_star}
    \sigma^* \simeq K(1 - 2 H_B^*)^{2/3} \quad \mbox{ for } H_B^* < 1/2,
\end{equation}
(and corresponds to decaying modes otherwise),
which is equivalent to
\begin{equation}
\label{eq:HG19_sigma}
\hat \sigma \simeq 0.42\left(\frac{1}{2} - \frac{H_B}{\hat{w}_f^2} \right)^{2/3} \hat{w}_f \hat{l}_f.
\end{equation}
This shows that $\hat \sigma$ is a monotonically increasing function of $\hat w_f$ and a monotonically decreasing function of $H_B$, and confirms that a vertical magnetic field can stabilize the parasitic shear instability. As a result, the finger velocity $\hat w_f$ is allowed to grow to larger amplitude in the presence of a magnetic field before the shear instability growth rate $\hat \sigma$ reaches the saturation condition Eq.~\eqref{eq:ingredient2}. This in turn leads to an increase in the magnitude of $\hat{F}_C$ according to Eq.~\eqref{eq:ingredient1}.

HG19 compared their model predictions for the compositional flux $\hat F_C$ (obtained by  
solving Eqs.~\eqref{eq:ingredient2} and \eqref{eq:HG19_sigma} numerically to obtain $\hat{w_f}$ as a function of $\hat \lambda_f$, and then substituting the result into Eq.~\eqref{eq:ingredient1}) to the values of $\hat F_C$ extracted from their DNS. They found that the two agree very well, at least for the limited set of parameters considered ($\Pr=\tau=D_B=0.1$, $R_0 = 1.45$). 
In what follows, we now extend their work to a much wider range of parameter space, and present new DNS of magnetized fingering convection that explore the effects of increasing the density ratio $R_0$ (to study less-unstable systems) and reducing the magnetic Prandtl number $\Pm$ (which reduces the dynamical connection between the field and the flow). Both extensions are important and relevant for stars, where $R_0$ necessarily becomes large near the boundary between a fingering-unstable region and a stable radiation zone \citep{Baltimore_paper}, and $\Pm$ is typically smaller than 1 in stellar radiative zones.

\section{Simulation results}
\label{sec:DNS_results}
\subsection{Methodology} \label{DNS:subsec:methods}

Our simulations are performed using the pseudospectral code PADDI \citep{Stellmach2011}, extended to solve equations \eqref{eq:dimless-mom}-\eqref{eq:dimless-divB} as detailed in \citet{Harrington_thesis} and used by HG19. 
Simulations are initialized with a uniform vertical magnetic field of unit strength, $\hat{\mathbf{B}} = (0, 0, 1)$, and all other dynamical fields set to zero. Low-amplitude white noise perturbations are added to $\hat{T}$ and $\hat{C}$ to seed the instability.

For all new simulations presented here, the horizontal dimensions of the simulation domain are taken as $L_x = L_y = 100$, which is adequate to accommodate 
many finger widths. 
However, the domain height $L_z$ varies across simulations to ensure that it is always substantially larger than the natural finger height \textcolor{black}{(as determined by visual inspection of simulation snapshots). In the majority of cases with extreme finger heights} (larger $H_B$ and $R_0$, as described in Sec.~\ref{DNS:subsec:R0} below), we performed additional simulations doubling $L_z$ to ensure that the saturated fluxes only change minimally in the process. 
We use domain heights as low as $L_z = 100$ and as large as $L_z = 800$ in the most extreme cases. 
\textcolor{black}{Further details on the effects of $L_z$ in such simulations are given in Appendix A of \citet{Traxler2011a}.}
Cubic simulations with $L_z = 100$ use a numerical resolution of $128^3$ Fourier modes (with standard dealiasing according to the 3/2 rule, so that nonlinearities are calculated in physical space on a $192^3$ grid)\footnote{All $R_0 = 1.45$ MHD simulations presented in this paper are taken from HG19, who used $192^3$ Fourier modes, with nonlinearities calculated in physical space on a $288^3$ grid.}, and simulations with taller domains use correspondingly larger numbers of Fourier modes in $z$ to ensure a fixed resolution. 
We check that this resolution is appropriate for all simulations by inspecting fluctuating fields and verifying that grid-scale oscillations (characteristic of the Gibbs phenomenon) are minimal. 

As expected from the linear theory of Sec.~\ref{setup:subsec:linear}, we find in all cases that the fingering instability  undergoes an exponential growth phase that eventually saturates into a quasi-stationary state of homogeneous fingering convection (see, e.g., HG19, Fig.~2). We extract the compositional flux by taking a time-average of $\hat F_C$ over this quasi-stationary state, and a measure of the error is taken to be the rms variability of the flux about this average; note, however, that in all figures presented below, the error bars are smaller than the symbol sizes themselves, and are thus entirely obscured.

\subsection{Trends with increased \texorpdfstring{$R_0$}{R0}} \label{DNS:subsec:R0}

\begin{figure}
    \plotone{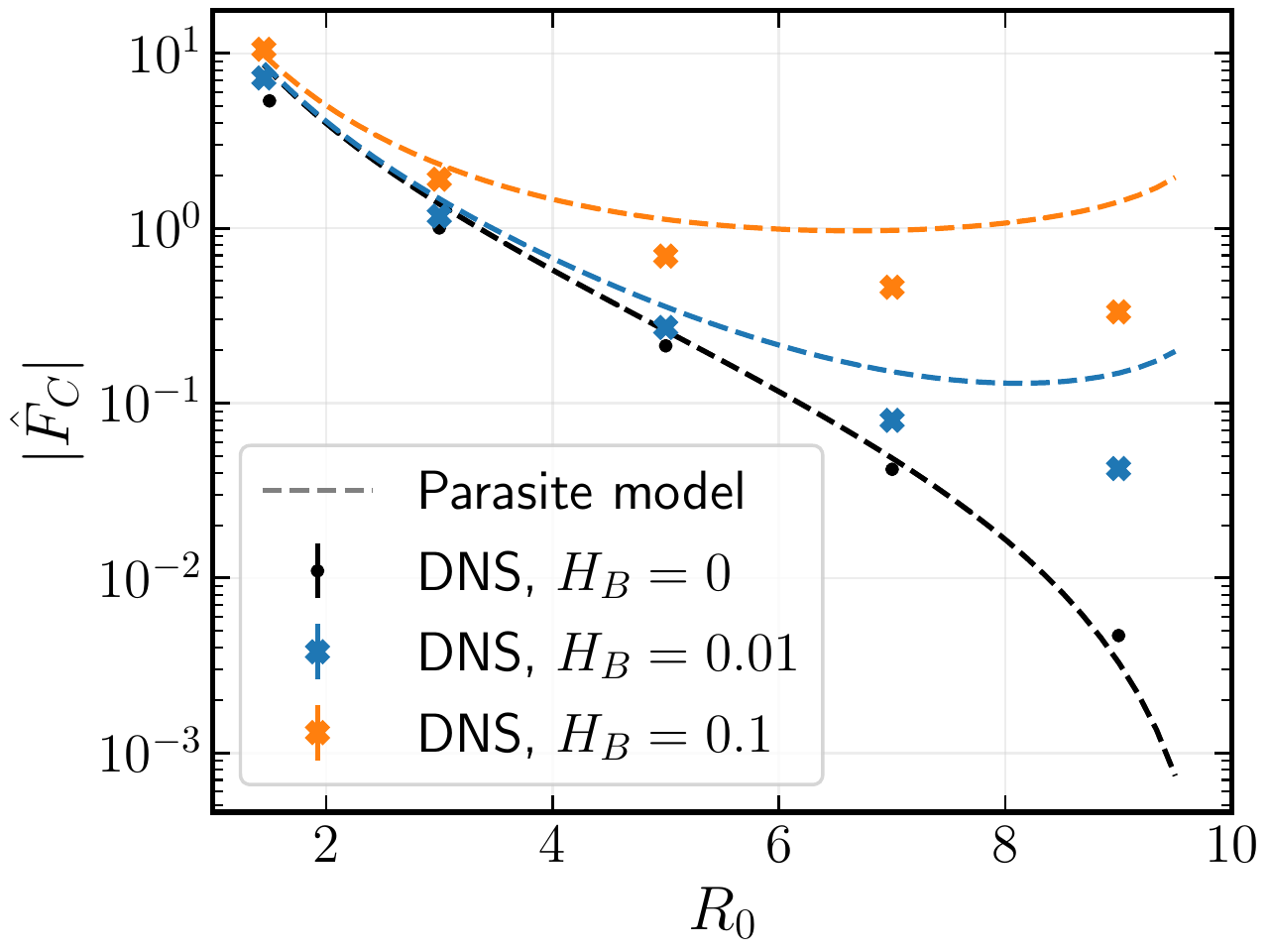}
    \caption{Turbulent compositional flux $|\hat{F}_C|$ as a function of density ratio $R_0$, for $\Pr = \tau = 0.1$ and $\Pm = 1$. Shown are results from direct numerical simulations without magnetic fields (black points), and in MHD with two different field strengths (blue and orange crosses, see legend), and compared in each case against the corresponding parasitic saturation model (dashed curve) described in Sec.~\ref{summary:subsec:parasites}.     
    At larger $R_0$ the fingering instability is weakest, and turbulent transport likewise becomes weaker as well. 
    The effect of increased magnetic field strength also becomes much more pronounced at large $R_0$, though this effect is over-predicted by parasitic saturation models.}
    \label{fig:FC_R0_HBscan_DNS}
\end{figure}

We start by discussing the effect of increasing the density ratio $R_0$, while holding the other physical parameters fixed as $\Pr = \tau = 0.1$ and $\Pm = 1$. Figure \ref{fig:FC_R0_HBscan_DNS} shows how $|\hat{F}_C|$ varies with $R_0$ for the hydrodynamic case (black points) and in MHD for two different values of $H_B$ (colored symbols), with the parasitic saturation models described in Sec.~\ref{summary:subsec:parasites} shown as dashed lines for comparison. 

For fixed $H_B$, we see that $|\hat{F}_C|$ decreases as $R_0 \to 1/\tau$. 
This is consistent with expectations from linear theory \citep[][HG19]{baines_gill_1969}, which shows that the fingering instability is gradually stabilized as $R_0$ increases. We also see that $|\hat{F}_C|$ increases with $H_B$ at fixed $R_0$, consistent with the findings of HG19. 
Our new simulations reveal that this effect is much more pronounced at larger $R_0$, where $|\hat{F}_C|$ can increase by more than one order of magnitude compared with the hydrodynamic case, even for moderate $H_B$. This can be explained at least qualitatively by the theory proposed by HG19 (see the dashed lines in Fig.~\ref{fig:FC_R0_HBscan_DNS})\textcolor{black}{, through the suppression of the parasitic mode by the magnetic field}, although the latter clearly overestimates the effect. 

\begin{figure*}
    \centering
    \includegraphics[width=\textwidth]{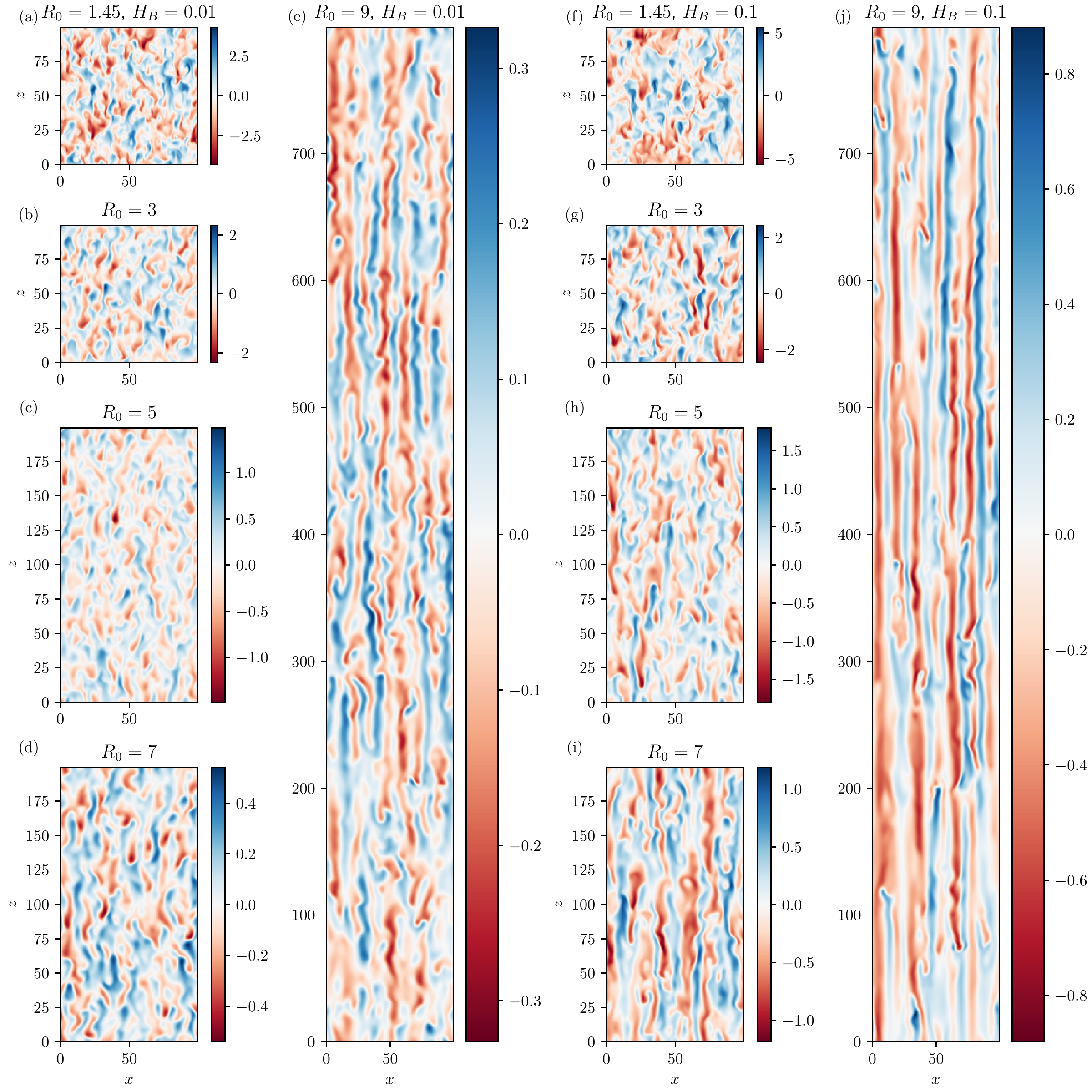}
    \caption{Snapshots of $\hat u_z$ taken during the statistically stationary state for DNS with increasing $R_0$ (see labels), at $\Pr = \tau = 0.1$, $D_B = 0.1$ ($\Pm=1$). Panels (a)-(e) show data for $H_B = 0.01$ and panels (f)-(j) show data for $H_B = 0.1$. We see that increasing $H_B$ increases the finger height, and the effect is stronger at high $R_0$ than at small $R_0$.}
    \label{fig:uz_snapshots_Pm1}
\end{figure*}

The enhanced effect of magnetic fields on the turbulence at large $R_0$ relative to small $R_0$ can also be seen qualitatively in Fig.~\ref{fig:uz_snapshots_Pm1}, which shows snapshots of $\hat{u}_z$ in the $x$-$z$ plane for the MHD simulations presented in Fig.~\ref{fig:FC_R0_HBscan_DNS}.
At $R_0 = 1.45$, we see that the velocity fields in the $H_B = 0.01$ and $H_B = 0.1$ simulations 
 have similar structures.
At larger $R_0$, however, the fingers are significantly longer and have higher flow speeds at larger $H_B$ than at smaller $H_B$\textcolor{black}{, consistent with the findings of Fig.~\ref{fig:FC_R0_HBscan_DNS} (see above)}. 

\subsection{Trends with decreased \texorpdfstring{$\Pm$}{Pm}} \label{DNS:subsec:Pm}

\begin{figure}
    \centering
    \includegraphics[width=\columnwidth]{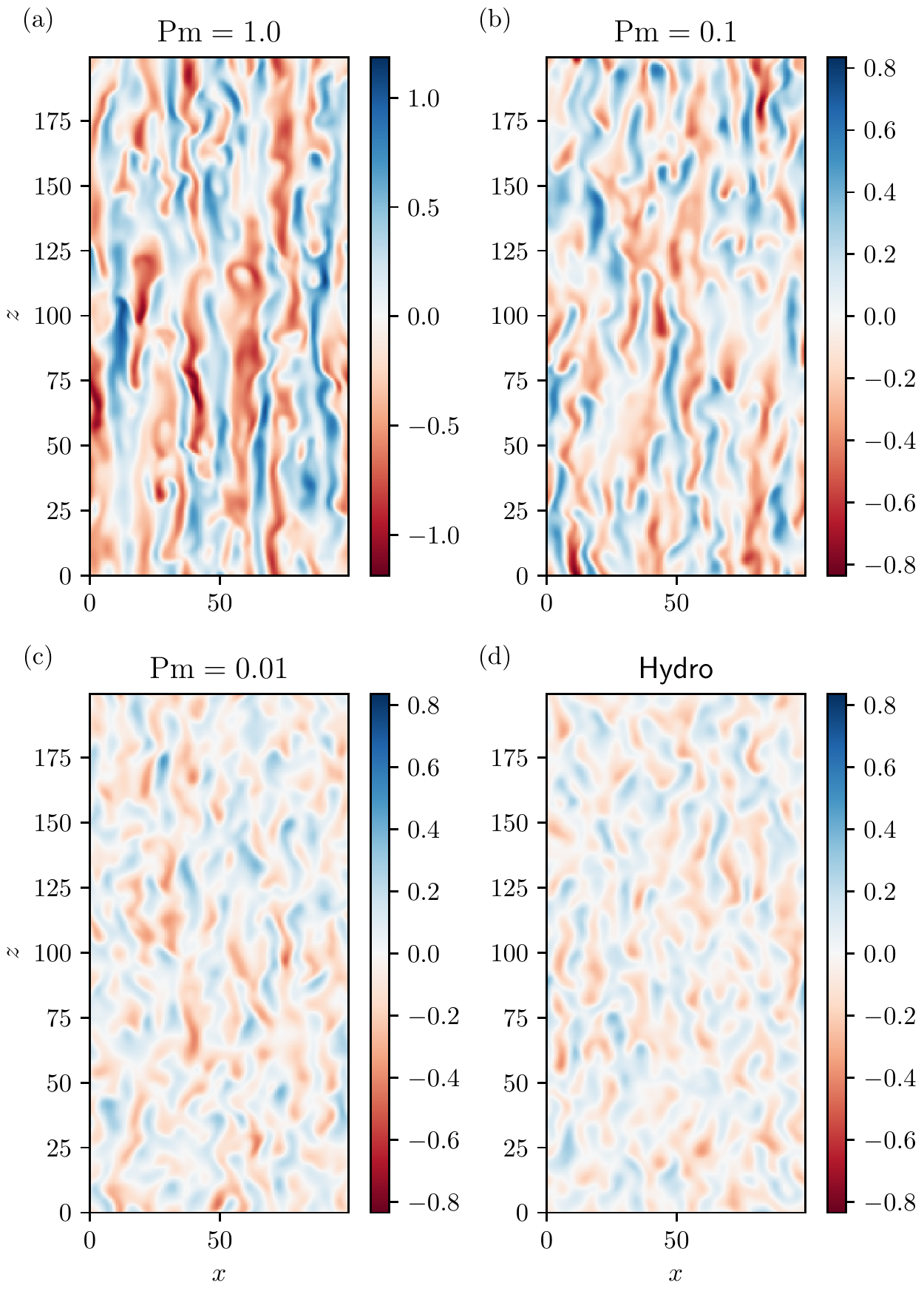}
    \caption{Snapshots of $\hat{u}_z$ as in Fig.~\ref{fig:uz_snapshots_Pm1}, with $\mathrm{Pr} = \tau = 0.1$ and $R_0 = 7$. 
    Panels (a)-(c) show magnetized simulations with $H_B = 0.1$ and $\Pm$ decreasing from $1$ to $0.01$, while (d) shows a hydrodynamic simulation for comparison. 
    Note that as $\Pm$ decreases, the tendency for magnetic fields to lengthen fingers becomes less pronounced, with the $\Pm = 0.01$ case largely indistinguishable from the hydrodynamic case. 
    }
    \label{fig:uz_snapshots_Pmscan}
\end{figure}

In stellar interiors, the magnetic diffusivity is typically somewhat larger than the kinematic viscosity, so the magnetic Prandtl number is roughly $\Pm \sim 0.01 - 1$. Meanwhile HG19 only presented simulations for $\Pm = 1$. 
Yet, it is well-known that MHD turbulence in $\Pm < 1$ fluids behaves very differently than at $\Pm = 1$ \citep[see, e.g.,][for prominent examples in the stellar dynamo context]{christensen_2006,warnecke_2023}. 
Thus, we now investigate the impact of using a smaller value of $\Pm$. 

The impact of reducing $\Pm$ can be seen qualitatively in Fig.~\ref{fig:uz_snapshots_Pmscan}, which shows snapshots of $\hat{u}_z$ as in Fig.~\ref{fig:uz_snapshots_Pm1}, with $\Pr = \tau = 0.1$, $R_0 = 7$, and, in panels (a) - (c), $H_B = 0.1$ with $\Pm$ decreasing from $1$ to $0.01$. 
Panel (d) shows a hydrodynamic simulation for comparison. 
We see that reducing $\Pm$ leads to shorter, lower-amplitude fingers, similar to the effect of reducing $H_B$ shown in Fig.~\ref{fig:uz_snapshots_Pm1}. 
At $\Pm = 0.01$, the dynamics are nearly indistinguishable from the hydrodynamic case.

\begin{figure*}
    \includegraphics[width=1.0\textwidth]{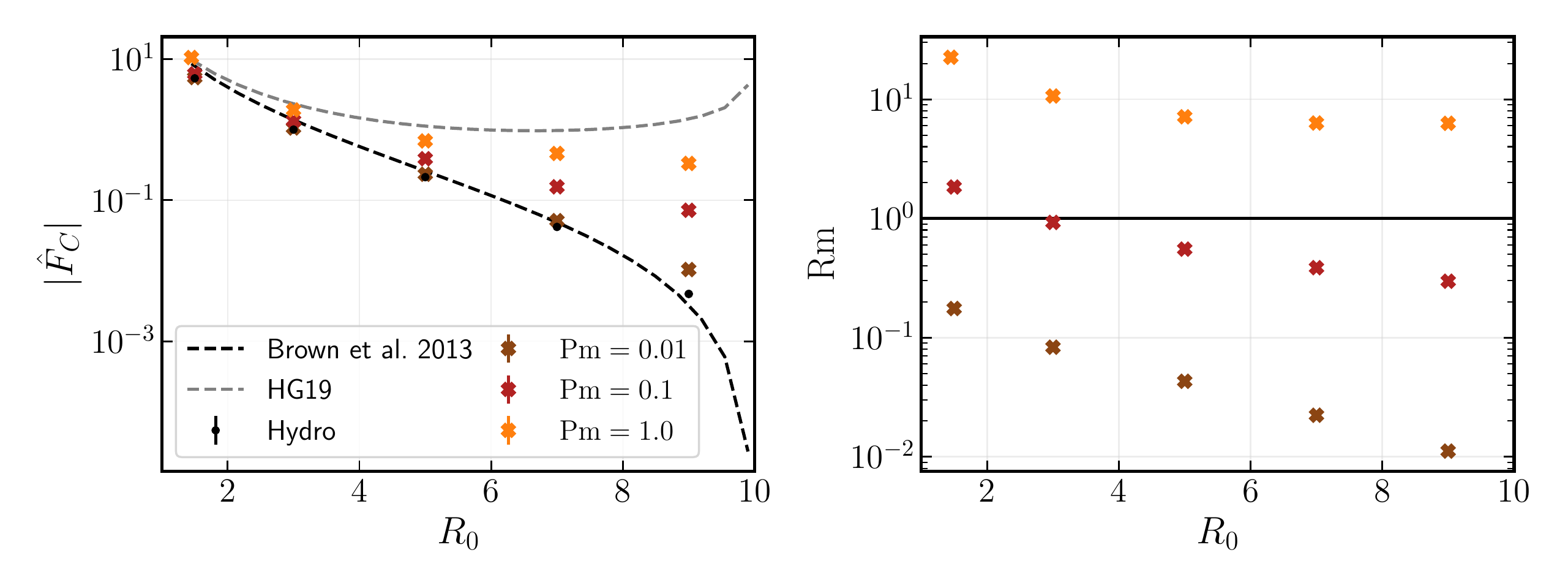}
    \caption{Left: Turbulent compositional flux $|\hat F_C|$ plotted against density ratio $R_0$ for the hydrodynamic case (black dots) and for $H_B = 0.1$ with different values of $\mathrm{Pm}$ (colored symbols, see legend for detail). All simulations use $\Pr = \tau = 0.1$. Decreasing $\Pm$ dramatically reduces $|\hat F_C|$, and the $\Pm = 0.01$ MHD simulations yield nearly identical fluxes to corresponding hydrodynamic simulations. The black dashed line shows the predictions from the \citet{Brown_2013} model, while the grey dashed line shows the predictions from the HG19 model; see text for detail.
    Right: an estimate of the effective magnetic Reynolds number $\mathrm{Rm}$ of the fingers for the MHD simulations in the left panel, estimated using \eqref{eq:rm_star}. The black horizontal line represents $\mathrm{Rm} = 1$.
    }
    \label{fig:FC_R0_Pmscan_DNS}
\end{figure*}

To look at this effect more quantitatively, we show in the left panel of Fig.~\ref{fig:FC_R0_Pmscan_DNS} the turbulent compositional flux against the density ratio $R_0$ (as in Fig.~\ref{fig:FC_R0_HBscan_DNS}) for simulations with $\Pr = \tau = 0.1$, for both hydrodynamic simulations (black dots) and MHD simulations (colored crosses) with fixed $H_B = 0.1$ and $\Pm$ varying from $1$ to $0.01$. 
Consistent with Fig.~\ref{fig:uz_snapshots_Pmscan}, we see that $|\hat{F}_C|$ decreases as $\Pm$ decreases, with the fluxes in $\Pm = 0.01$ simulations nearly identical to the hydrodynamic case, except for the largest value of $R_0$. 
The predicted fluxes from both hydrodynamic (black, \citet{Brown_2013}) and MHD (grey, HG19) parasitic saturation models are also shown as dashed curves for comparison with the data.
Because HG19 ignored viscosity and resistivity in their calculation of $\hat{\sigma}$ (see Sec.~\ref{summary:subsec:parasites}), their model does not depend on $\Pm$, which is why there is only one corresponding dashed curve. 
It is therefore clear that their model, by construction, cannot capture the strong dependence of $\hat{F}_C$ on $\Pm$ observed in our DNS data, especially at large $R_0$. 

One may wonder why the HG19 model can successfully explain the properties of fingering convection at their selected parameter values ($\Pr = \tau =0.1$, $\Pm=1$ and $R_0 =1.45$), while dramatically failing in our new higher $R_0$, lower $\Pm$  simulations. 
One plausible explanation for this breakdown in the model is provided in the right panel of Fig.~\ref{fig:FC_R0_Pmscan_DNS}, where the data points represent one possible measure of the magnetic Reynolds number $\mathrm{Rm}$ associated with the vertical velocity of the fingers, for each of the simulations presented on the left panel. 
We define this magnetic Reynolds number in terms of the rms vertical velocity $\uzrms$ and the wavenumber of the fastest-growing elevator modes $\hat{l}_f$, as 
\begin{equation}\label{eq:rm_star}
    \mathrm{Rm} \equiv \frac{\uzrms}{D_B \hat{l}_f} =  \frac{\Pm \uzrms}{\Pr \hat{l}_f}.
\end{equation}
We find that while $\mathrm{Rm}$ is greater than one in the $\Pm = 1$ simulations, the $\Pm = 0.1$ and $\Pm = 0.01$ simulations all have $\mathrm{Rm} \lesssim 1$. 
Thus, it is no longer appropriate to ignore the effects of resistivity (and in some cases, even of viscosity) in the calculation of the shear instability growth rate $\hat{\sigma}$, which may explain why the HG19 model fails at these parameters. 
In what follows, we now extend the HG19 model to account for 
\textcolor{black}{additional effects including dissipation and buoyancy} 
in the computation of the parasitic modes.

\section{Extending the parasitic saturation model}
\label{sec:extending_HG19}
\subsection{\textcolor{black}{The extended model}}
\label{parasite_models:subsec:model}
\textcolor{black}{Here we extend} 
the HG19 model to take into account 
\textcolor{black}{diffusive effects (including viscosity and resistivity) as well as temperature and composition fluctuations} 
on the development of the parasitic instability. 
\textcolor{black}{As we will see, including diffusion alone is insufficient to bring the model into agreement with simulations, and one must additionally include temperature ($T$) and composition ($C$) fluctuations as was done by \citet{RadkoSmith2012}.}

As in \citet{Brown_2013} and HG19, we use the frozen-in approximation to assume that the fastest-growing elevator mode has a two-dimensional, sinusoidal, steady velocity field of the form $\hat{\mathbf{u}}_0 = \hat{w}_f \cos(\hat{l}_f x) \hat{\mathbf{e}}_z$ (the choice of using $x$ as the oscillating direction is made here without loss of generality). It is embedded in a uniform, streamwise magnetic field $\hat{\mathbf{B}}_0 = \hat{\mathbf{e}}_z$. For sufficiently large values of $\hat{w}_f$, the elevator mode becomes unstable to a parasitic instability whose growth rate $\hat{\sigma}$ can be computed by performing a standard linear stability calculation. \textcolor{black}{Ignoring buoyancy fluctuations, this calculation would be the same as 
the stability calculation that was performed in \citet{FraserJFM22}, albeit with a different non-dimensionalization.} 

\citet{FraserJFM22} found that 2D perturbations are always the most unstable \textcolor{black}{in their system}. 
\textcolor{black}{We thus assume that this is also true here, and consequently, that} the fastest-growing parasitic modes can be described using a streamfunction formulation where the perturbed flow and field are given by $\mathbf{\hat{u}} = \nabla \times \left( \hat{\psi} \hat{\mathbf{e}}_y \right)$ and $\mathbf{\hat{b}} = \nabla \times \left( \hat{A} \hat{\mathbf{e}}_y \right)$, respectively. 
We split the streamfunction $\hat{\psi}$ into elevator and perturbation as follows: $\hat{\psi} = \hat{\psi}_E + \hat{\psi}_P$ (where the subscript $E$ represents the elevator mode and $P$ the perturbation)\textcolor{black}{, and similarly let $\hat T = \hat T_E + \hat T_P$ and $\hat C = \hat C_E + \hat C_P$}. Since the elevator mode does not affect the background field, we have $\hat{A} = \hat{A}_P$.
As the elevator mode is invariant in $z$ and sinusoidal in $x$, we express it as:
\begin{equation}
\label{eq:psiEdef}
    \hat \psi_E = -i \hat{E}_{\psi}\left( e^{ix\hat{l}_f} - e^{-ix \hat{l}_f} \right). 
\end{equation}
The elevator mode amplitude ($\hat{E}_{\psi}$) is related to $\hat{w}_f$ by the equation $2\hat{E}_{\psi}\hat{l}_f = \hat{w}_f$, where $\hat l_f$ is the horizontal wavenumber of the fastest-growing fingering mode for given values of $\Pr$, $\tau$ and $R_0$ ($\hat{l}_f$ can easily be computed from linear theory, see section \ref{setup:subsec:linear}).
\textcolor{black}{The $\hat{T}$ and $\hat{C}$ components of the elevator mode are similarly given by (cf.~Eq~[3.10] of \citealt{RadkoSmith2012})}
\textcolor{black}{
\begin{equation}
    \hat{T}_{E} = \hat{E}_{T}\left(e^{ix\hat{l}_f} + e^{-ix\hat{l}_f}\right)
\end{equation}
and
\begin{equation}
    \hat{C}_{E} = \hat{E}_{C}\left(e^{ix\hat{l}_f} + e^{-ix\hat{l}_f}\right),
\end{equation}
where $\hat{E}_T$ and $\hat{E}_C$ are related to $\hat E_{\psi}$ (and therefore $\hat w_f$) via
\begin{equation}
\hat{E}_T = \frac{- \hat{l}_f \hat{E}_{\psi} } { \hat{\lambda}_f  + \hat{l}_f^2 }, \quad \hat{E}_C = \frac{ - \hat{l}_f \hat{E}_{\psi} }{ R_0 \left(\hat{\lambda}_f +\tau  \hat{l}_f^2\right) }. 
\end{equation}}

On 
\textcolor{black}{their} 
frozen-in spatially sinusoidal background state, \citet{FraserJFM22} found that the fastest growing linear perturbations have the same spatial period as the elevator mode. 
\textcolor{black}{Here we assume the same, thus perturbations take the form}
\textcolor{black}{
\begin{equation}\label{eq: fourier modes_TC}
	\begin{pmatrix}
		\hat{\psi}_{P} \\ \hat{T}_{P} \\ \hat{C}_{P} \\ \hat{A}_P
	\end{pmatrix} = 
	e^{\hat{\sigma} t + i \hat{k}_z z}
    \sum_{m = -N}^{N} \begin{pmatrix}
						\hat{\psi}_m \\ \hat{T}_m \\ \hat{C}_m \\ \hat{A}_m
					  \end{pmatrix}  e^{i  m\hat{l}_f x },
\end{equation}}
where $\hat k_z$ is the vertical wavenumber of the perturbation, and $\hat \sigma$ is its growth rate. 
In practice, we truncate the series to a finite sum from $-N$ to $N$ so that we can numerically compute $\hat{\sigma}$ for a given $\hat k_z$. 
Using 
\eqref{eq: fourier modes_TC} in the governing equations (\ref{eq:dimless-mom})-(\ref{eq:dimless-divB}), 
linearizing the system (assuming perturbations have small amplitudes compared with the background state), and projecting the results onto each Fourier mode yields the following set of linear equations:
\textcolor{black}{
\begin{multline}\label{eq: shear_linear_start_TC}
	-\hat{\sigma} \hat{k}_m^2 \hat{\psi}_m -
	  i \hat{l}_f  \hat{k}_z   \hat{E}_{\psi} \left( \hat{k}_{m+1}^2 \hat{\psi}_{m+1} +  \hat{k}_{m-1}^2 \hat{\psi}_{m-1} \right)  +  \\
 i \hat{l}_f^3 \hat{k}_z \hat{E}_{\psi}\left( \hat{\psi}_{m+1} + \hat{\psi}_{m-1}\right)  \\ 
 =  \text{Pr} \hat{k}_m^4 \hat \psi_m + i\text{Pr} m\hat{l}_f(\hat{T}_m -\hat{C}_m) -   
	i H_B  \hat{k}_z \hat{k}_m^2 \hat{A}_m , 
 \end{multline}
\begin{multline}
	\hat{\sigma} \hat{T}_m  +  i \hat{l}_f  \hat{k}_z \hat{E}_{\psi}(\hat{T}_{m+1} + \hat{T}_{m-1})  + \hat{l}_f \hat{k}_z \hat{E}_{T} (\hat{\psi}_{m-1} -\hat{\psi}_{m+1} ) \\ + im  \hat{l}_f \hat{\psi}_m = -\hat{k}_m^2 \hat{T}_m, 
\end{multline}
\begin{multline}
    \hat{\sigma} \hat{C}_m  +  i \hat{l}_f  \hat{k}_z \hat{E}_{\psi} (\hat{C}_{m+1} + \hat{C}_{m-1})  + \hat{l}_f \hat{k}_z \hat{E}_{C} (\hat{\psi}_{m-1} -\hat{\psi}_{m+1} )
 \\ + im \hat{l}_f \frac{\hat{\psi}_m }{R_0}= -\tau \hat{k}_m^2 \hat{C}_m, 
\end{multline}
and 
\begin{multline}\label{eq: shear_linear_end_TC}
    \hat{\sigma} \hat{A}_m +    i  \hat{l}_f \hat{k}_z \hat{E}_{\psi}\left(  \hat{A}_{m+1} +  \hat{A}_{m-1} \right) 
   \\ = -D_B \hat{k}_m^2 \hat{A}_m + i \hat{k}_z  \hat{\psi}_m,
\end{multline}}
where we have defined $\hat{k}_m^2 = m^2 \hat{l}_f ^2 + \hat{k}_z^2$. For each given values of $\hat{k}_z$ and $\hat{w}_f$ and input parameters ($\mathrm{Pr}$, $\tau$, $R_0$, $H_B$, $D_B$), Eqs.~\eqref{eq: shear_linear_start_TC} - \eqref{eq: shear_linear_end_TC} define an eigenvalue problem with possibly multiple eigenvalues $\hat{\sigma}$. 
For each $\hat{k}_z$, we only consider the eigenvalue with the largest real part and call this $\hat \sigma(\hat k_z,\hat w_f; \Pr, \tau, R_0, H_B, D_B)$. We then maximize the result over all $\hat{k}_z$ to obtain the fastest-growing parasitic mode growth rate as a function of the elevator mode velocity and input parameters, $\hat \sigma_{\rm max}(\hat w_f; \Pr, \tau, R_0, H_B, D_B)$.  
Finally, knowing the elevator mode's growth rate $\hat{\lambda}_f$ (which is again a known function of $\Pr$, $\tau$ and $R_0$) we find the smallest value of $\hat{w}_f$ for which Eq.~\eqref{eq:ingredient2} holds (corresponding to the largest amplitude the elevator mode can reach before being disrupted by the parasitic mode), and use the result to calculate $\hat{F}_C$ using Eq.~\eqref{eq:ingredient1}. \textcolor{black}{For validation, we have checked (in the hydrodynamic case) that this procedure exactly recovers Figure 7 of \cite{RadkoSmith2012} with the parameters $\Pr = 7$, $\tau = 0.01$ (setting $C_2 = 1/4.3$ and $C_1 = 0.5$ to match their model parameters).}

As in the HG19 model, the entire procedure depends on two parameters $C_1$ and $C_2$ which appear in equations \eqref{eq:ingredient1} and \eqref{eq:ingredient2} respectively.
For now, we choose $C_1 = 0.62$ and  $C_2 = 0.33$ as our fiducial parameters 
\textcolor{black}{(we revisit this matter in Sec.~\ref{parasite_models:subsec:model_parameters})}.

\begin{figure*}
    \includegraphics[width=1.0\textwidth]{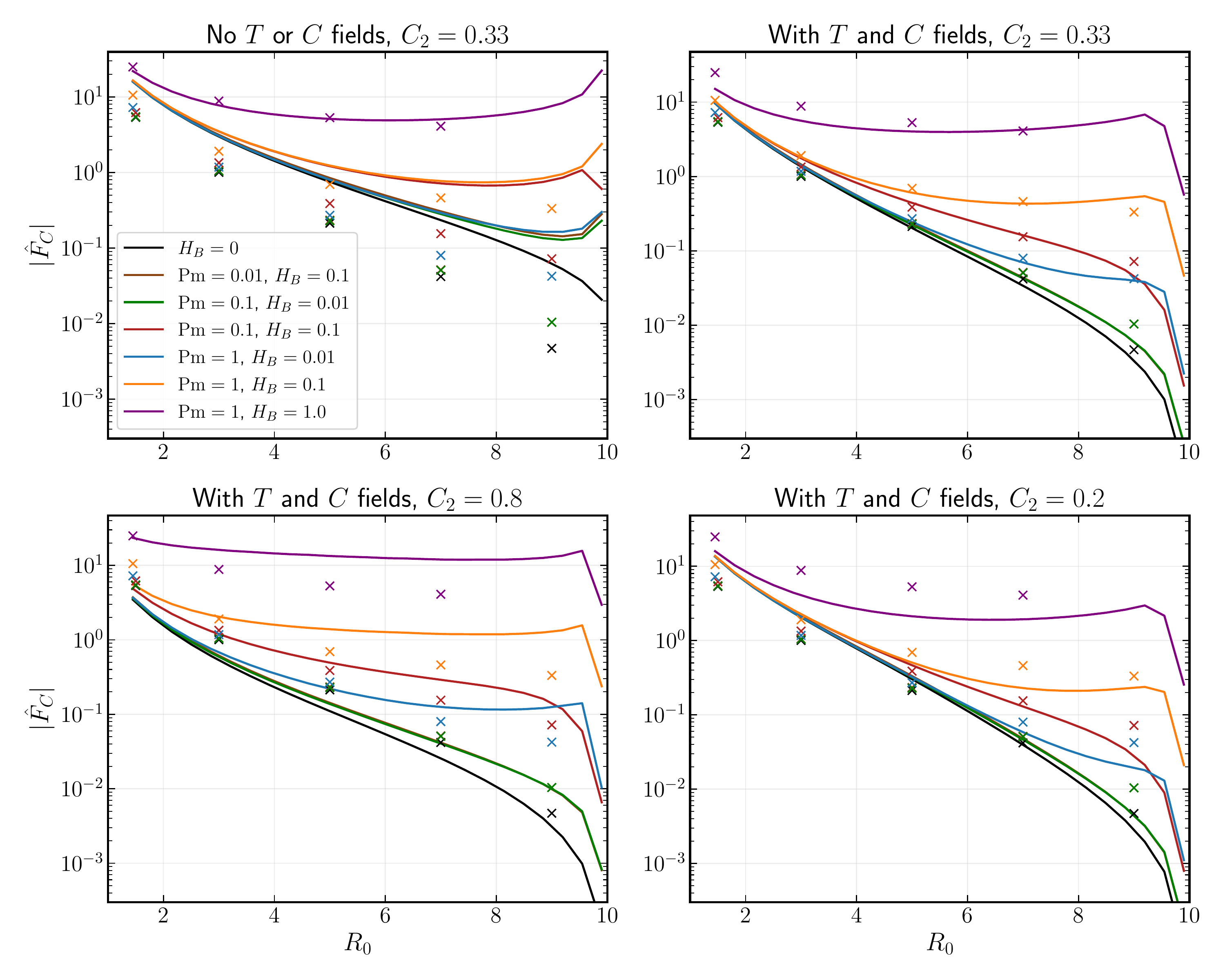}
    \caption{Turbulent compositional flux $|\hat{F}_C|$ as a function of density ratio $R_0$ for $\mathrm{Pr} = \tau = 0.1$ 
    \textcolor{black}{and multiple values of $\mathrm{Pm}$ and $H_B$ (see legend; note that the $H_B=0.1$, $\mathrm{Pm}=0.01$ and the $H_B=0.01$, $\mathrm{Pm}=0.1$ cases essentially overlap)}. The dissipative parasitic model (solid lines) is compared to the DNS results (crosses) when the temperature and composition fields are ignored (upper left) and when they are included (other panels). The value used for the parameter $C_2$ is shown in the title, and the corresponding value of $C_1$ is given in Table \ref{tab:error}.} 
    \label{fig:full_DNS_comparison}
\end{figure*}

The model predictions are compared with the data \textcolor{black}{for a range of $H_B$, $\Pm$, and $R_0$ values and with $\Pr = \tau = 0.1$} in 
\textcolor{black}{Fig.~\ref{fig:full_DNS_comparison}, where the top-left panel shows the dissipative model without $T$ and $C$ fluctuations included (cf.~\citealt{FraserJFM22}), and the top-right panel includes the full system described by Eqs.~\eqref{eq: shear_linear_start_TC}-\eqref{eq: shear_linear_end_TC}. In both panels, dissipative models are given by solid lines and DNS data by crosses.} 
\textcolor{black}{The number of Fourier modes used for this figure is 9 for the results without $\hat T$ and $\hat C$ (corresponding to $N=4$) and 21 when $\hat T$ and $\hat C$ are included (corresponding to $N=10$).
These numbers were chosen so that adding additional modes did not change the results significantly.}
\textcolor{black}{The top-left panel shows} 
that adding dissipation does not, on its own, help resolve the discrepancy between model and DNS. \textcolor{black}{Comparing to Fig.~\ref{fig:FC_R0_Pmscan_DNS}}, 
we see that adding viscosity significantly {\it worsens} the fit between the model and the data at large $R_0$ \textcolor{black}{\emph{even in the hydrodynamic case}}. 
This is because the predicted value of $\hat{w}_f$ would normally decreases as $R_0$ approaches $1/\tau$ in the non-dissipative model (and in the DNS data), but must exceed a certain critical value for the parasitic modes to overcome viscous damping in the dissipative model, and is hence significantly overestimated. For this reason, the model prediction for $|\hat{F}_C|$ largely overestimates the DNS results in that limit. 

In the magnetic case 
the problem is similar (with $|\hat F_C|$ being vastly over-estimated at large $R_0$), although the dissipative model is somewhat better at capturing the data trends than the non-dissipative HG19 model. In particular, we see that the dissipative model correctly predicts that $|\hat{F}_C|$ should decrease as $\Pm$ decreases, but the predicted sensitivity of $\hat{F}_C$ to changes in $\Pm$ is much weaker than what is seen in DNS, particularly at large $R_0$.

\textcolor{black}{The results of the full model with $T$ and $C$ fluctuations included is given in the top-right panel of Fig.~\ref{fig:full_DNS_comparison}.} 
\textcolor{black}{Focusing first on the hydrodynamic case (black),} we see 
that including temperature and composition in the parasitic mode calculation greatly improves the accuracy of the prediction for $\hat F_C$ over that of the dissipative model without them, to the extent that it now recovers the quality of the original non-dissipative model of \citet{Brown_2013}\textcolor{black}{---see Fig.~\ref{fig:FC_R0_Pmscan_DNS}}. But more importantly, we also see 
that the new model is far better at capturing the dependence of $\hat F_C$ on 
\textcolor{black}{$H_B$}, $R_0$, and $\Pm$ in the magnetic case. In particular, we now correctly predict that the turbulent compositional flux at $H_B = 0.1$ tends to its corresponding hydrodynamic value when $\Pm \rightarrow 0$, which is expected as the magnetic field has an increasingly small effect on the flow in this limit.


In hindsight, we can see why including the $\hat T$ and $\hat C$ fields might be conceptually important as follows. If we ignore $\hat T$ and $\hat C$ in the dynamics of the parasitic mode, shear is only possible source of instability, and is easily stabilized by viscosity or magnetic fields when the elevator mode amplitude $\hat{w}_f$ is small. For this reason, $\hat w_f$ must always exceed a certain threshold for the parasitic mode to grow, which leads us to  overestimate $|\hat F_C|$ in some regions of parameter space. 
However, when $\hat T$ and $\hat C$ are included, low-$\hat k_z$ parasitic modes can draw energy from the unstable background compositional stratification just like the basic fingering instability does, and can therefore be unstable even at very low (or even zero) $\hat w_f$. This seemed to have already been known by \citet{RadkoSmith2012} who had included the $\hat T$ and $\hat C$ fields in their original study of the parasitic modes and their role in saturating the hydrodynamic fingering instability in the geophysical context. 

\subsection{Model parameters} \label{parasite_models:subsec:model_parameters}
As mentioned before, the model presented has two free parameters, which are assumed to be universal constants. The first, $C_1$,  appears in Eq.~\eqref{eq:ingredient1} and controls the level of chemical mixing driven by elevator modes of a given amplitude $\hat{w}_f$. Adjusting $C_1$ simply corresponds to uniformly shifting the curves shown in 
Fig.~\ref{fig:full_DNS_comparison} vertically up or down. The second parameter, $C_2$, represents the assumed ratio between the growth rate of the elevator mode and its parasitic perturbation at saturation (see equation Eq.~\ref{eq:ingredient2}). Its effect on the prediction for $\hat F_C$ is more subtle, as shown in Fig.~\ref{fig:full_DNS_comparison} (bottom row). 

Two new values of $C_2$ are tested: 
$C_2 = 0.8$ (lower left) and $C_2 = 0.2$ (lower right). In both cases, $C_1$ is adjusted \textit{a posteriori} to obtain the best fit with the data, see Table \ref{tab:error} for detail. This figure shows that, generally speaking, a larger value of $C_2$ causes the model to be more sensitive to increases in $H_B$ and  $\Pm$. This can be understood fairly intuitively as follows. For small values of $C_2$, the parasitic mode growth rate $\hat \sigma$ needs to reach a correspondingly larger value before saturation can occur, which therefore requires a larger amplitude $\hat{w}_f$. In turn, this implies that the effective magnetic Reynolds number of the parasitic mode is larger, reducing the effects of the magnetic diffusivity. Similarly, the elevator mode kinetic energy is higher, so the ambient magnetic energy is less significant in comparison, and the field has less impact on the parasitic perturbations.
For larger values of $C_2$ the opposite is true. Comparing the three figures, we find that $C_2 = 0.33$ is roughly optimal because it has the right sensitivity to both $H_B$ and $\Pm$. 

In practice, we have chosen the values $C_1 = 0.62$ and $C_2= 0.33$ as our fiducial parameters using the following procedure. 
We define the rms error between the model and the data as follows:
\begin{equation}
\label{eq:error}
    \text{E} = \sqrt{\frac{1}{N_p}\sum_{R_0, \Pm,H_B} \left(  \log(|\hat{F}_{C,\text{DNS}}|) - \log(|\hat{F}_{C,\text{model}}|)  \right)^2 } ,
\end{equation}
where the sum ranges over the available DNS data (which has $\Pr = \tau = 0.1$) and $N_p$ is the number of data points. Note that we use the logarithm of the flux to ensure that all data points carry a similar weight. For a given value of $C_2$, we first compute the model flux $\hat{F}_{C,\text{model}}^{*}$ assuming that $C^*_1 = 1$. Then the multiplicative factor $C_1$ that best fits the data for that given $C_2$ satisfies
\begin{equation}
\log(C_1) = \frac{1}{N_p}\sum_{R_0, \Pm,H_B} \left(  \log(|\hat{F}_{C,\text{DNS}}|) - \log(|\hat{F}_{C,\text{model}}^{*}|)  \right). 
\end{equation}
Finally, we compute $\text{E}$ using \eqref{eq:error}. 
 The results are shown in Table \ref{tab:error} for a range of $C_2$. We can see among all of the values of $C_2$ tested, the choice $C_2 = 0.33, C_1 = 0.62$ minimizes the rms error $\text{E}$.

\begin{table}
\begin{center}
\begin{tabular}{ccc}
   \hline
    $C_2$ & $C_1$ & $\text{E}$ \\
    \hline
     
    1.0 & 34.03 & 0.670\\ 
0.9 & 5.14 & 0.632\\ 
0.8 & 3.29 & 0.601\\ 
0.66 & 2.06 & 0.503\\ 
0.5 & 1.23 & 0.339\\
{\bf 0.33} & {\bf 0.62} & {\bf 0.308}\\ 
0.25 & 0.39 & 0.453\\ 
0.2 & 0.27 & 0.590\\ 
0.1 & 0.08 & 0.988\\ \hline
\end{tabular}
\caption{Rms error E between the model and the data, computed using (\ref{eq:error}), for different values of $C_2$. The associated value of $C_1$ (see text for details) is also shown. Chosen fiducial parameters are marked in bold.}
\label{tab:error}
\end{center}
\end{table}

\begin{figure*}
    \centering
    \includegraphics[width=\linewidth]{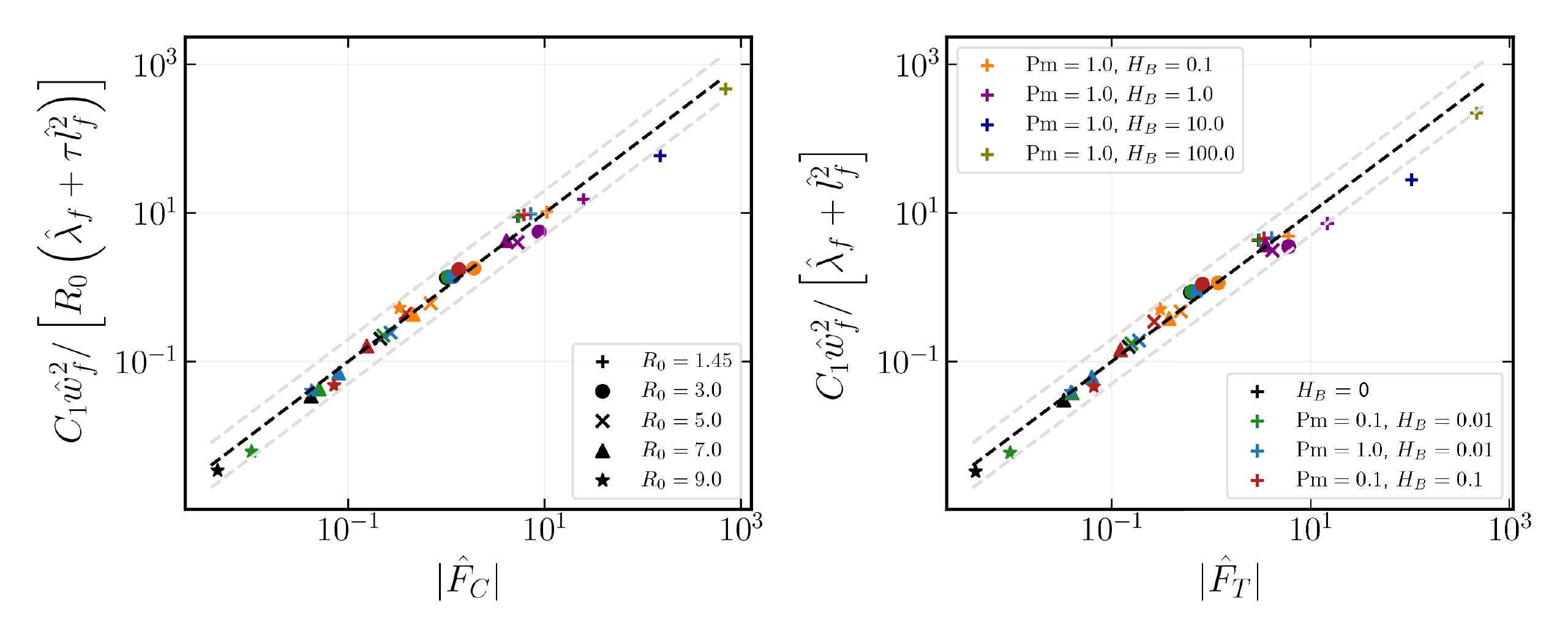}
    \caption{Left: Turbulent compositional flux $|\hat{F}_C|$ obtained from DNS, plotted against parasitic model prediction (see Section \ref{parasite_models:subsec:model} for detail). Right: The same for the turbulent thermal flux $|\hat{F}_T|$. The model parameters used in this figure are $C_2 = 0.33$, $C_1 = 0.62$.
    The black dashed line corresponds to equality of the model and data and the gray dashed lines  
    correspond to a factor of two discrepancy between the model and data. The marker shape indicates the value of $R_0$ (as shown in the legend in the left plot) and marker color indicates the values of $H_B$ and $\Pm$ (as shown in the legend in the right plot). All parameters are also shown in Fig.~\ref{fig:full_DNS_comparison}  
    with the addition of two points corresponding to $H_B = 10$ and $H_B = 100$ from HG19.
    }
    \label{fig:final_scatter}
\end{figure*}

Finally, note that although $C_2$ and $C_1$ were tuned using DNS data for the compositional flux $\hat{F}_C$ only, the model is equally accurate in predicting the thermal flux $\hat{F}_T$ with the same model parameters, where $\hat{F}_T$ is calculated from $\hat{w}_f$ according to 
\begin{equation}
    \hat{F}_T = -C_1 \frac{\hat{w}_f^2}{\hat{\lambda}_f + \hat{l}_f^2},
\end{equation}
just as Eq.~\eqref{eq:ingredient1} is used to calculate $\hat{F}_C$ \citep{RadkoSmith2012,Brown_2013}.
This is shown in Figure \ref{fig:final_scatter}, which compares the model and the data over a wide range of input parameters for both $\hat{F}_C$ and $\hat{F}_T$, using the tuned parameters ($C_2 = 0.33, C_1 = 0.62$). We see that the model is accurate at predicting both fluxes to within a factor of two (i.e. within the grey lines) for almost all cases.

\section{Summary and Conclusions}
\label{sec:conclusions}
\subsection{Summary}
 \label{sec:ccl_summary}

 We have performed a suite of DNS of magnetized fingering convection at $\Pr= \tau = 0.1$ with an initially uniform, vertical magnetic field, and compared the measured turbulent compositional fluxes to predictions from the parasitic saturation model put forth by HG19. Our work extends HG19's simulation study (which used the same $\Pr$ and $\tau$, and fixed the density ratio at $R_0 = 1.45$, and the magnetic Prandtl number at $\Pm = 1$) to higher $R_0$ and lower $\Pm$, which are both relevant to stellar interior conditions. 

As in HG19, we found that the large-scale vertical magnetic field always enhances vertical transport by fingering convection. This is especially true at large $R_0$, where it elongates fingers and increases transport even more than at small $R_0$. However, while predictions of the simple parasitic saturation model proposed by HG19 are qualitatively consistent with this trend, we have found that they significantly over-estimate the magnitude of the turbulent fluxes measured in our DNS at large $R_0$. Lowering $\Pm$ has the opposite effect: in our simulations with $\Pm < 1$, the effect of magnetic fields on the fingers becomes less pronounced and the turbulent transport is correspondingly lower than when $\Pm = 1$. This is because the resistivity causes 
the magnetic field to decouple from the fluid, at which point the turbulent flow behaves essentially as in the corresponding hydrodynamic simulations. Because the HG19 model neglects viscosity and resistivity in the calculation of the parasitic shear instability growth rate $\hat{\sigma}$, its predictions do not depend on $\Pm$, so it is unable, by construction, to capture this trend. 

By computing the magnetic Reynolds number for the fingers, $\Rm$, we found that the HG19 model fails to quantitatively predict the correct turbulent fluxes whenever $\Rm$ is order unity or smaller. Thus, magnetic diffusion is important to the dynamics of the fingers and their parasitic shear instabilities and cannot be neglected (at least in the parameter regimes covered by the DNS). 

In an effort to remedy the problem, we 
extended the HG19 model to include \textcolor{black}{(a)} viscosity and resistivity\textcolor{black}{, and (b) temperature and composition fluctuations} in the calculation of the growth rates of the parasitic modes. 
\textcolor{black}{While these effects were unnecessary to explain the data in \citet{Brown_2013} and HG19 and thus deemed negligible, we found that \emph{both} of these additions are necessary to reproduce the data when considering} the wider range of parameter space explored in this paper.
Accounting for \textcolor{black}{these effects} in the parasitic saturation model 
now provides an excellent match with the turbulent thermal and compositional flux data for all available simulations (see Fig.~\ref{fig:final_scatter}). We therefore conclude that the model presented in Section \ref{parasite_models:subsec:model} can reliably be used to estimate mixing by magnetized fingering convection in the presence of a uniform, vertical magnetic field for a very wide range of possible input parameters ($R_0,\Pr,\tau,D_B$ and $H_B$). 

\subsection{Model caveats}
\label{sec:ccl_caveats}

There are several caveats one must bear in mind, however. First, a \textcolor{black}{uniform,} vertical magnetic field is a very special geometric configuration\textcolor{black}{. While the uniform field assumption is justified by the extremely small-scale nature of the fingers compared to the pressure scale height \citep[and thus presumably the gradient scale lengths of whatever magnetic fields exist in these stellar radiation zones; see][Table 1]{Harrington_2019},} 
\textcolor{black}{fingers are in general more likely to encounter oblique fields than vertical ones.}
HG19 demonstrated that the fastest-growing mode of the fingering instability is unchanged if the field is inclined, but that the nonlinear saturation process is affected. More specifically, they saw in several examples that the turbulent mixing in inclined field cases can greatly exceed that of the corresponding vertical field configuration \citep[see][for examples with a horizontal field]{Harrington_thesis}. If this result holds in general, it implies that the parasitic saturation model proposed in this paper only provides a {\it lower bound} to the magnetized fingering flux amplitudes. Future work will therefore be needed to quantify the effects of field inclination on 
fingering convection. 

Second, great care must be taken in the application of the parasitic saturation model to ensure that, for any given set of input parameters, the predictions respect the constraints of the Boussinesq approximation. More specifically, one must check that the dimensional elevator mode velocity at saturation, namely $w_f = \hat w_f \kappa_T/d$, always remains much smaller than the local sound speed, and that the characteristic vertical scale of the fingers  remains much smaller than the local pressure, density or temperature scale height. Both of these self-consistency checks can easily be done \textit{a posteriori}. Note that a rough estimate of finger height is $d \hat k_{z,{\rm max}}^{-1}$, where  $\hat k_{z,{\rm max}}^{-1}$ is the nondimensional vertical wavenumber of the fastest-growing parasitic mode at saturation.

Finally, we note that the computation of the fingering fluxes in our complete magnetized parasitic saturation model (including buoyancy and diffusion) is computationally much more onerous than in \citet{Brown_2013} or HG19. As such, it is not entirely practical for stellar evolution calculations that need to be performed for a large number of stars over their entire lifetimes. However, it can in principle be implemented in \textcolor{black}{such} codes 
\textcolor{black}{\citep[work is currently under way on an implementation in MESA,][]{MESA2013}},
with the caveat that it will greatly slow down the computations. 
Possible ways of accelerating the flux calculations are discussed in Section \ref{sec:ccl_models}.

\subsection{Implications for stellar abundance observations}
\label{sec:ccl_obs}

\begin{figure*}
\includegraphics[width=\textwidth]{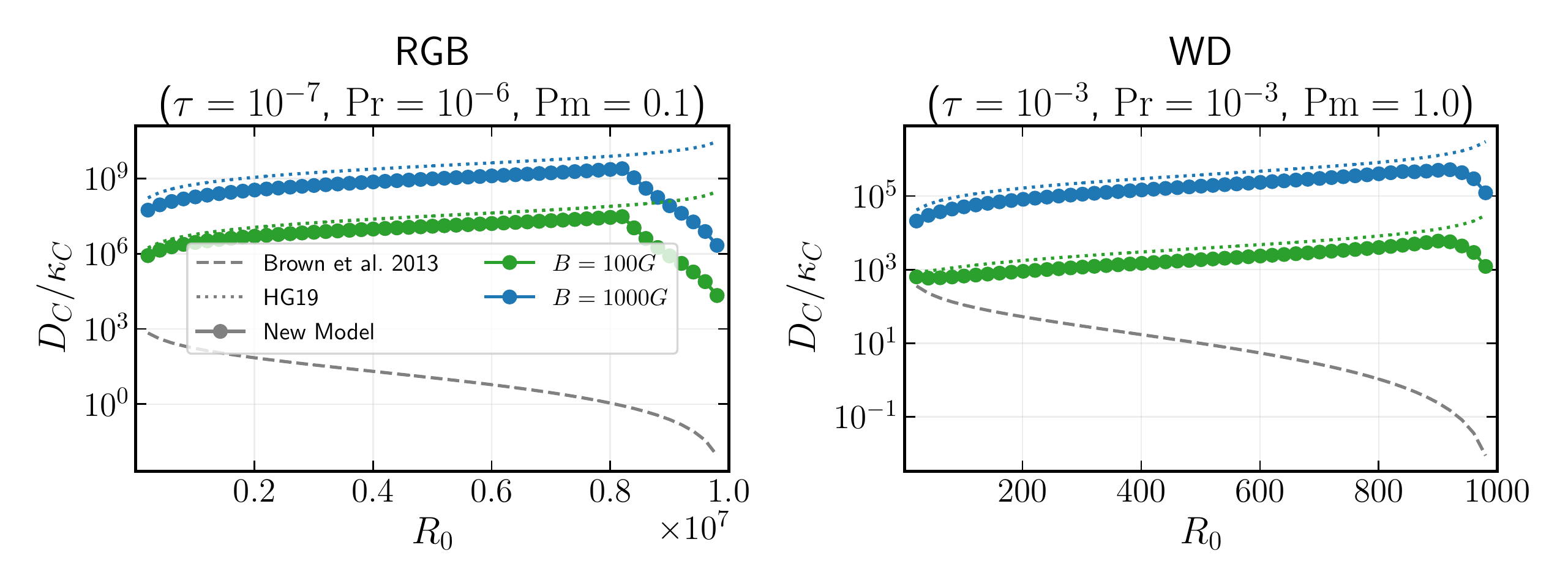}
    \caption{Turbulent compositional diffusivity $D_C$ (see equation \ref{eq:DCdef}) normalized by microscopic diffusivity $\kappa_C$, predicted by various models, for parameter values appropriate of the fingering region of a typical RGB star (left) and a typical DA white dwarf (right), see main text for detail. In each panel, the dashed line corresponds to the \citet{Brown_2013} hydrodynamic model, and the colored dotted lines correspond to the HG19 model. Finally, the round symbols correspond to the new parasitic model of Section \ref{parasite_models:subsec:model} The case for $B_0 = 100$G is shown in green and corresponds to $H_B = 10^{-7}$ for the RGB star, and $H_B = 10^{-3}$ for the white dwarf (cf. HG19). The case for $B_0=1000$G correspond to $H_B = 10^{-5}$ for the RGB star, and $H_B = 10^{-1}$ for the white dwarf.
    }
    \label{fig:stellar}
\end{figure*}

With these caveats in mind, we can now apply the complete parasitic saturation model described in Section \ref{parasite_models:subsec:model} to two situations where fingering convection is known to be important in stars, as introduced in Section \ref{sec:intro}: just above the hydrogen-burning shell of RGB stars, and just below the surface of white dwarfs undergoing accretion from a companion or a debris disk. Using the stellar models for a typical 1$M_\odot$ RGB star just before the luminosity bump, and a typical 0.59$M_\odot$, $T_{\rm eff} = 11,150$K DA white dwarf \citep[both models are presented in][]{Garaud2015}, we can obtain very rough estimates\footnote{Note that these quantities can vary by one or two orders of magnitude in the fingering regions of RGB stars and white dwarfs, so the numbers provided here are indicative only.} for the fingering region viscosity, thermal diffusivity, and magnetic diffusivity. The former are computed as described in \citet{Garaud2015}, and the latter is computed using the standard \citet{Spitzer1962} formula for non-degenerate gases, which is valid in these regions. We find that for RGB stars typical parameter values are $\Pr = O(10^{-6})$, $\tau = O(10^{-7})$, $\Pm = O(0.1)$, while for the near-surface layers of white dwarfs, $\Pr \simeq \tau = O(10^{-3})$, $\Pm = O(1)$. Figure \ref{fig:stellar} shows the corresponding model predictions for $D_C/\kappa_C$, where $D_C$ is the turbulent mixing coefficient for composition resulting from the fingering convection (see equation \ref{eq:DCdef}). The left panel shows the RGB star model, while the right panel shows the white dwarf model. In both cases, we present two values of $H_B$ corresponding to a 100G field (green dots) and a 1000G field (blue dots), respectively (see caption for detail). We also show, for reference, the model predictions from HG19 as dotted lines of the same colors, and the hydrodynamic model of \citet{Brown_2013} as a dashed line. We have confirmed \textit{a posteriori} that in all cases, the model predictions are well within the region validity of the Boussinesq approximation (see caveats described in Sec.~\ref{sec:ccl_caveats}). 

We immediately see that for these parameter values, a magnetic field of 100G-1000G is sufficient to increase mixing by fingering convection by many orders of magnitude above the levels expected for the purely hydrodynamic instability alone, especially in the RGB star. This finding is consistent with the conclusions of HG19, and demonstrates that even a relatively weak magnetic field can have an enormous impact on vertical transport by this instability. As a result, we conclude that fingering convection is indeed sufficient to explain the rapid heavy-element surface abundance changes observed in RGB stars around the luminosity bump \citep{Gratton}, as originally proposed by \citet{CharbonnelZahn2007}, {\it as long as a weak magnetic field is also present.} In fact, the tight correlation between $D_C$ and $H_B$ seen in Fig. \ref{fig:stellar} (where $D_C \propto H_B$ depends quadratically on the field strength over much of the fingering range) might be used to constrain the interior field strength in the vicinity of the fingering region using the surface abundance observations \citep[cf.][]{Baltimore_paper}. 

With regards to the white dwarf results, we also find that vertical transport by the fingering instability is greatly enhanced by the field, especially at larger density ratios. This implies that the downward transport by fingering convection of the heavy elements accreted onto the surface from a debris disk \citep[cf.][]{Deal_WD,bauer_polluted_2019} will be much faster in strongly magnetized white dwarfs than in non-magnetized or weakly magnetic ones. As a result, one may expect to see an inverse correlation between the magnetic field strength and the surface abundance of heavy elements in white dwarfs undergoing active accretion. Equivalently, this implies that a much larger accretion rate than previously thought is required to explain surface abundances of heavy elements in magnetic white dwarfs than in non-magnetic ones, as already noted by HG19. 

\subsection{Implications for stellar evolution models}
\label{sec:ccl_models}

Figure \ref{fig:stellar} also shows that the HG19 model (dotted lines) and the new complete parasitic saturation model (large dots) are consistent with one another (except at very large $R_0$, see below) within the margin of error of the model\footnote{Indeed, the factor of about 2 difference in the low and moderate $R_0$ regime is not unexpected given that the new model is itself only accurate within a factor of 2 of the DNS data.}. 
This can easily be explained \textit{a posteriori} by computing the estimated magnetic Reynolds number of the fingers $\Rm$ (using equation \eqref{eq:rm_star} and replacing $\hat u_{z,{\rm rms}}$ with $\hat w_f$) and noting that it is much larger than one at these parameter values, except when 
$R_0 \rightarrow \tau^{-1}$ .
Hence, {\it the major discrepancies reported in Section \ref{sec:DNS_results} between the DNS data and the HG19 model can be understood, in retrospect, as an artefact of using much larger values of $\Pr$ and $\tau$ in the DNS than in real stars}. When $\Pr$ and $\tau$ are much smaller, the predicted magnetic Reynolds number $\mathrm{Rm}$ of the fingers is correspondingly much larger, and HG19 generally applies except at very large $R_0$.

 The complete parasitic model and the HG19 model do differ significantly in the limit $R_0 \rightarrow \tau^{-1}$, however, \textcolor{black}{despite the fact that $\mathrm{Rm}$ remains large for most of that range} 
 \textcolor{black}{(for instance, for the $B = 1000$G case for the RGB parameters at $R_0 = 8 \times 10^6$, where the HG19 model and the new model begin to disagree significantly, the new model predicts $\mathrm{Rm} \approx 700$)}. Instead, in that limit the parasitic modes take the form of 
 \textcolor{black}{buoyancy-driven} modes whose growth rate is amplified by the shear, which were absent in HG19 because their model neglected buoyancy. As a result, HG19 failed to capture the decrease in the amount of turbulent mixing that must inevitably occur as the system becomes stable to fingering convection. The complete parasitic model, on the other hand, correctly predicts that $D_C \rightarrow 0$ in that limit. As such, we strongly advocate for its use over HG19 if at all possible.
 
Yet, as mentioned earlier, the added computational cost is quite substantial, and a simplified parametrization of the new model would be preferable. To create it, one may leverage the knowledge that HG19 is adequate when $\Rm \gg 1$ {\it and} $R_0 \ll \tau^{-1}$, and use it when these conditions are satisfied. The complete model would then only be needed to compute the turbulent fluxes when $\Rm = O(1)$ or less, and/or $R_0 \rightarrow \tau^{-1}$. 
\textcolor{black}{Work towards implementing such a model in MESA is currently in progress.}

\subsection{Other model predictions}
 \label{sec:ccl_cool}

There are several other interesting implications of the new model that are worth mentioning as well. First, because the magnetic Reynolds number $\Rm$ is much greater than one in most of the fingering range at stellar values of $\Pr$, $\tau$ and $\Pm$, it is quite plausible that the fluid motions associated with the saturated fingering convection could drive a small-scale dynamo. 
Whether that small-scale dynamo would then substantially change the model predictions, however, remains to be determined. 

Another interesting prospect is that because the magnetic field can greatly enhance the turbulent temperature and compositional fluxes in fingering regions of stars, they may now become unstable to large-scale mean-field instabilities such as the collective instability and the layering instability \citep[see the review by][]{garaud_DDC_review}. \citet{Garaud2015} had shown that the turbulent fingering fluxes are too weak to trigger these instabilities in the hydrodynamic case. Our results suggest that they may actually be active in the magnetized case (although will necessarily also be affected by the large-scale field). This will be the subject of future work. 

In the meantime, this study has once more illustrated the power of parasitic saturation models in predicting the nonlinear saturation of a particular instability and the resulting turbulent mixing it causes. The key, as we have demonstrated here, is to take great care in ensuring that {\it all} physical processes are included, so as to capture all possible parasitic modes in the calculation -- or otherwise run the risk of vastly overestimating or underestimating the level of saturation. Based on their natural similarities, we anticipate that a similar parasitic model could adequately predict the saturation of the magnetized GSF instability, for instance, as well as many other instabilities where saturation occurs because of parasitic modes.  

\subsection*{Acknowledgements}
AF and PG acknowledge support from National Science Foundation grant AST-1908338. 
AF acknowledges additional support from NASA HTMS grant 80NSSC20K1280, and from the George Ellery Hale Postdoctoral Fellowship in Solar, Stellar and Space Physics at University of Colorado. 
Simulations performed using the Extreme Science and Engineering Discovery Environment (XSEDE) Expanse supercomputer at the San Diego Supercomputer Center through allocation TG-PHY210050, and the Lux supercomputer at UC Santa Cruz, funded by NSF MRI grant AST-1828315.
We gratefully acknowledge helpful discussions with Matteo Cantiello and Daniel Lecoanet at the KITP program "Probes of transport in stars" (supported in part by the National Science Foundation under Grant No.~NSF PHY-1748958), 
helpful discussions with Justin Walker, David Hughes, Timour Radko, Stephan Stellmach, and Rafa Fuentes\textcolor{black}{, and helpful discussions with Evan Bauer and Rich Townsend who, while working on a separate project, independently verified some of the calculations and figures in this work}.


\software{Matplotlib \citep{Matplotlib}, NumPy \citep{numpy}, SciPy \citep{2020SciPy-NMeth}}

\bibliographystyle{aasjournal}
\bibliography{Fraser_Garaud_MHD_DDC.bib}

\begin{thebibliography}{}
\expandafter\ifx\csname natexlab\endcsname\relax\def\natexlab#1{#1}\fi
\providecommand{\url}[1]{\href{#1}{#1}}
\providecommand{\dodoi}[1]{doi:~\href{http://doi.org/#1}{\nolinkurl{#1}}}
\providecommand{\doeprint}[1]{\href{http://ascl.net/#1}{\nolinkurl{http://ascl.net/#1}}}
\providecommand{\doarXiv}[1]{\href{https://arxiv.org/abs/#1}{\nolinkurl{https://arxiv.org/abs/#1}}}

\bibitem[{Baines \& Gill(1969)}]{baines_gill_1969}
Baines, P.~G., \& Gill, A.~E. 1969, Journal of Fluid Mechanics, 37, 289,
  \dodoi{10.1017/S0022112069000553}

\bibitem[{Barker {et~al.}(2019)Barker, Jones, \& Tobias}]{barker2019}
Barker, A.~J., Jones, C.~A., \& Tobias, S.~M. 2019, Monthly Notices of the
  Royal Astronomical Society, 487, 1777, \dodoi{10.1093/mnras/stz1386}

\bibitem[{Barker {et~al.}(2020)Barker, Jones, \& Tobias}]{barker2020}
---. 2020, Monthly Notices of the Royal Astronomical Society, 495, 1468,
  \dodoi{10.1093/mnras/staa1327}

\bibitem[{Bauer \& Bildsten(2018)}]{bauer_increases_2018}
Bauer, E.~B., \& Bildsten, L. 2018, \apj, 859, L19,
  \dodoi{10.3847/2041-8213/aac492}

\bibitem[{Bauer \& Bildsten(2019)}]{bauer_polluted_2019}
---. 2019, \apj, 872, 96, \dodoi{10.3847/1538-4357/ab0028}

\bibitem[{{Brown} {et~al.}(2013){Brown}, {Garaud}, \& {Stellmach}}]{Brown_2013}
{Brown}, J.~M., {Garaud}, P., \& {Stellmach}, S. 2013, Astrophys. J., 768, 34,
  \dodoi{10.1088/0004-637X/768/1/34}

\bibitem[{{Charbonnel} \& {Zahn}(2007)}]{CharbonnelZahn2007}
{Charbonnel}, C., \& {Zahn}, J.-P. 2007, Astron. Astrophys., 467, L15,
  \dodoi{10.1051/0004-6361:20077274}

\bibitem[{Christensen \& Aubert(2006)}]{christensen_2006}
Christensen, U.~R., \& Aubert, J. 2006, Geophysical Journal International, 166,
  97, \dodoi{10.1111/j.1365-246X.2006.03009.x}

\bibitem[{{Deal} {et~al.}(2013){Deal}, {Deheuvels}, {Vauclair}, {Vauclair}, \&
  {Wachlin}}]{Deal_WD}
{Deal}, M., {Deheuvels}, S., {Vauclair}, G., {Vauclair}, S., \& {Wachlin},
  F.~C. 2013, \aap, 557, L12, \dodoi{10.1051/0004-6361/201322206}

\bibitem[{{Denissenkov}(2010)}]{Denissenkov}
{Denissenkov}, P.~A. 2010, \apj, 723, 563, \dodoi{10.1088/0004-637X/723/1/563}

\bibitem[{{Eddington}(1926)}]{Eddington}
{Eddington}, A.~S. 1926, {The Internal Constitution of the Stars} (Cambridge
  University Press)

\bibitem[{Farihi {et~al.}(2009)Farihi, Jura, \&
  Zuckerman}]{farihi_infrared_2009}
Farihi, J., Jura, M., \& Zuckerman, B. 2009, The Astrophysical Journal, 694,
  805, \dodoi{10.1088/0004-637X/694/2/805}

\bibitem[{Fraser {et~al.}(2022{\natexlab{a}})Fraser, Cresswell, \&
  Garaud}]{FraserJFM22}
Fraser, A.~E., Cresswell, I.~G., \& Garaud, P. 2022{\natexlab{a}}, Journal of
  Fluid Mechanics, 949, A43, \dodoi{10.1017/jfm.2022.782}

\bibitem[{Fraser {et~al.}(2022{\natexlab{b}})Fraser, Joyce, Anders, Tayar, \&
  Cantiello}]{Baltimore_paper}
Fraser, A.~E., Joyce, M., Anders, E.~H., Tayar, J., \& Cantiello, M.
  2022{\natexlab{b}}, The Astrophysical Journal, 941, 164,
  \dodoi{10.3847/1538-4357/aca024}

\bibitem[{Fuentes {et~al.}(2023)Fuentes, Cumming, Castro-Tapia, \&
  Anders}]{fuentes_heat_2023}
Fuentes, J.~R., Cumming, A., Castro-Tapia, M., \& Anders, E.~H. 2023, The
  Astrophysical Journal, 950, 73, \dodoi{10.3847/1538-4357/accb56}

\bibitem[{Garaud(2018)}]{garaud_DDC_review}
Garaud, P. 2018, Annual Review of Fluid Mechanics, 50, 275,
  \dodoi{10.1146/annurev-fluid-122316-045234}

\bibitem[{Garaud(2021)}]{garaud_journey_2021}
---. 2021, Physical Review Fluids, 6, 030501,
  \dodoi{10.1103/PhysRevFluids.6.030501}

\bibitem[{Garaud {et~al.}(2015)Garaud, Medrano, Brown, Mankovich, \&
  Moore}]{Garaud2015}
Garaud, P., Medrano, M., Brown, J.~M., Mankovich, C., \& Moore, K. 2015, The
  Astrophysical Journal, 808, 89, \dodoi{10.1088/0004-637X/808/1/89}

\bibitem[{Ginzburg {et~al.}(2022)Ginzburg, Fuller, Kawka, \&
  Caiazzo}]{ginzburg_slow_2022}
Ginzburg, S., Fuller, J., Kawka, A., \& Caiazzo, I. 2022, Monthly Notices of
  the Royal Astronomical Society, 514, 4111, \dodoi{10.1093/mnras/stac1363}

\bibitem[{{Goodman} \& {Xu}(1994)}]{Goodman_Xu}
{Goodman}, J., \& {Xu}, G. 1994, \apj, 432, 213, \dodoi{10.1086/174562}

\bibitem[{{Gratton} {et~al.}(2000){Gratton}, {Sneden}, {Carretta}, \&
  {Bragaglia}}]{Gratton}
{Gratton}, R.~G., {Sneden}, C., {Carretta}, E., \& {Bragaglia}, A. 2000, \aap,
  354, 169

\bibitem[{Harrington(2018)}]{Harrington_thesis}
Harrington, P. 2018, Master's thesis, UC Santa Cruz.
\newblock \url{https://escholarship.org/uc/item/5hw2v310}

\bibitem[{Harrington \& Garaud(2019)}]{Harrington_2019}
Harrington, P.~Z., \& Garaud, P. 2019, The Astrophysical Journal, 870, L5,
  \dodoi{10.3847/2041-8213/aaf812}

\bibitem[{Holyer(1984)}]{holyer_1984}
Holyer, J.~Y. 1984, Journal of Fluid Mechanics, 147, 169,
  \dodoi{10.1017/S0022112084002044}

\bibitem[{{Hunter}(2007)}]{Matplotlib}
{Hunter}, J.~D. 2007, Computing in Science and Engineering, 9, 90,
  \dodoi{10.1109/MCSE.2007.55}

\bibitem[{Isern {et~al.}(2017)Isern, García-Berro, Külebi, \&
  Lorén-Aguilar}]{isern_common_2017}
Isern, J., García-Berro, E., Külebi, B., \& Lorén-Aguilar, P. 2017, The
  Astrophysical Journal Letters, 836, L28, \dodoi{10.3847/2041-8213/aa5eae}

\bibitem[{Jura(2003)}]{jura_tidally_2003}
Jura, M. 2003, The Astrophysical Journal, 584, L91, \dodoi{10.1086/374036}

\bibitem[{{Kippenhahn} {et~al.}(1980){Kippenhahn}, {Ruschenplatt}, \&
  {Thomas}}]{kippenhahn_thermohaline}
{Kippenhahn}, R., {Ruschenplatt}, G., \& {Thomas}, H.~C. 1980, \aap, 91, 175

\bibitem[{{Koester} {et~al.}(2014){Koester}, {G{\"a}nsicke}, \&
  {Farihi}}]{Koester}
{Koester}, D., {G{\"a}nsicke}, B.~T., \& {Farihi}, J. 2014, \aap, 566, A34,
  \dodoi{10.1051/0004-6361/201423691}

\bibitem[{Latter {et~al.}(2009)Latter, Lesaffre, \& Balbus}]{latter_mri_2009}
Latter, H.~N., Lesaffre, P., \& Balbus, S.~A. 2009, Monthly Notices of the
  Royal Astronomical Society, 394, 715,
  \dodoi{10.1111/j.1365-2966.2009.14395.x}

\bibitem[{Longaretti \& Lesur(2010)}]{longaretti_mri-driven_2010}
Longaretti, P.-Y., \& Lesur, G. 2010, Astronomy \& Astrophysics, 516, A51,
  \dodoi{10.1051/0004-6361/201014093}

\bibitem[{{Mochkovitch}(1983)}]{Mochkovitch}
{Mochkovitch}, R. 1983, \aap, 122, 212

\bibitem[{{Montgomery} \& {Dunlap}(2023)}]{Montgomery}
{Montgomery}, M.~H., \& {Dunlap}, B.~H. 2023, arXiv e-prints, arXiv:2312.11647,
  \dodoi{10.48550/arXiv.2312.11647}

\bibitem[{{Paxton} {et~al.}(2013){Paxton}, {Cantiello}, {Arras}, {Bildsten},
  {Brown}, {Dotter}, {Mankovich}, {Montgomery}, {Stello}, {Timmes}, \&
  {Townsend}}]{MESA2013}
{Paxton}, B., {Cantiello}, M., {Arras}, P., {et~al.} 2013, \apjs, 208, 4,
  \dodoi{10.1088/0067-0049/208/1/4}

\bibitem[{Pessah(2010)}]{Pessah}
Pessah, M.~E. 2010, The Astrophysical Journal, 716, 1012,
  \dodoi{10.1088/0004-637X/716/2/1012}

\bibitem[{Pessah \& Goodman(2009)}]{Pessah_Goodman}
Pessah, M.~E., \& Goodman, J. 2009, The Astrophysical Journal, 698, L72,
  \dodoi{10.1088/0004-637X/698/1/L72}

\bibitem[{Radko(2013)}]{radko_book}
Radko, T. 2013, Double-Diffusive Convection (Cambridge University Press),
  \dodoi{10.1017/CBO9781139034173}

\bibitem[{{Radko} \& {Smith}(2012)}]{RadkoSmith2012}
{Radko}, T., \& {Smith}, D.~P. 2012, J. Fluid Mech., 692, 5,
  \dodoi{10.1017/jfm.2011.343}

\bibitem[{{Renzo} \& {G{\"o}tberg}(2021)}]{Renzo}
{Renzo}, M., \& {G{\"o}tberg}, Y. 2021, \apj, 923, 277,
  \dodoi{10.3847/1538-4357/ac29c5}

\bibitem[{{Salaris} \& {Cassisi}(2017)}]{Salaris_review}
{Salaris}, M., \& {Cassisi}, S. 2017, Royal Society Open Science, 4, 170192,
  \dodoi{10.1098/rsos.170192}

\bibitem[{Sengupta \& Garaud(2018)}]{sengupta_2018}
Sengupta, S., \& Garaud, P. 2018, The Astrophysical Journal, 862, 136,
  \dodoi{10.3847/1538-4357/aacbc8}

\bibitem[{Shetrone {et~al.}(2019)Shetrone, Tayar, Johnson, Somers,
  Pinsonneault, Holtzman, Hasselquist, Masseron, Mészáros, Jönsson, Hawkins,
  Sobeck, Zamora, \& García-Hernández}]{shetrone_constraining_2019}
Shetrone, M., Tayar, J., Johnson, J.~A., {et~al.} 2019, The Astrophysical
  Journal, 872, 137, \dodoi{10.3847/1538-4357/aaff66}

\bibitem[{Spiegel \& Veronis(1960)}]{spiegel_boussinesq_1960}
Spiegel, E.~A., \& Veronis, G. 1960, The Astrophysical Journal, 131, 442,
  \dodoi{10.1086/146849}

\bibitem[{{Spitzer}(1962)}]{Spitzer1962}
{Spitzer}, L. 1962, {Physics of Fully Ionized Gases}

\bibitem[{Stancliffe {et~al.}(2007)Stancliffe, Glebbeek, Izzard, \&
  Pols}]{stancliffe_carbon-enhanced_2007}
Stancliffe, R.~J., Glebbeek, E., Izzard, R.~G., \& Pols, O.~R. 2007, Astronomy
  \& Astrophysics, 464, L57, \dodoi{10.1051/0004-6361:20066891}

\bibitem[{{Stellmach} {et~al.}(2011){Stellmach}, {Traxler}, {Garaud},
  {Brummell}, \& {Radko}}]{Stellmach2011}
{Stellmach}, S., {Traxler}, A., {Garaud}, P., {Brummell}, N., \& {Radko}, T.
  2011, J. Fluid Mech., 677, 554, \dodoi{10.1017/jfm.2011.99}

\bibitem[{Stern(1960)}]{Stern_1960}
Stern, M. 1960, Tellus, 12, 172

\bibitem[{Stevenson(1980)}]{stevenson_eutectic_1980}
Stevenson, D.~J. 1980, Le Journal de Physique Colloques, 41, C2,
  \dodoi{10.1051/jphyscol:1980209}

\bibitem[{Tayar \& Joyce(2022)}]{tayar_joyce_2022}
Tayar, J., \& Joyce, M. 2022, The Astrophysical Journal Letters, 935, L30,
  \dodoi{10.3847/2041-8213/ac85ab}

\bibitem[{{Traxler} {et~al.}(2011){Traxler}, {Stellmach}, {Garaud}, {Radko}, \&
  {Brummell}}]{Traxler2011a}
{Traxler}, A., {Stellmach}, S., {Garaud}, P., {Radko}, T., \& {Brummell}, N.
  2011, J. Fluid Mech., 677, 530.
\newblock \doarXiv{1008.1807}

\bibitem[{Ulrich(1972)}]{ulrich_thermohaline_1972}
Ulrich, R.~K. 1972, The Astrophysical Journal, 172, 165, \dodoi{10.1086/151336}

\bibitem[{{van der Walt} {et~al.}(2011){van der Walt}, {Colbert}, \&
  {Varoquaux}}]{numpy}
{van der Walt}, S., {Colbert}, S.~C., \& {Varoquaux}, G. 2011, Computing in
  Science and Engineering, 13, 22, \dodoi{10.1109/MCSE.2011.37}

\bibitem[{{Virtanen} {et~al.}(2020){Virtanen}, {Gommers}, {Oliphant},
  {Haberland}, {Reddy}, {Cournapeau}, {Burovski}, {Peterson}, {Weckesser},
  {Bright}, {van der Walt}, {Brett}, {Wilson}, {Jarrod Millman}, {Mayorov},
  {Nelson}, {Jones}, {Kern}, {Larson}, {Carey}, {Polat}, {Feng}, {Moore}, {Vand
  erPlas}, {Laxalde}, {Perktold}, {Cimrman}, {Henriksen}, {Quintero}, {Harris},
  {Archibald}, {Ribeiro}, {Pedregosa}, {van Mulbregt}, \&
  {Contributors}}]{2020SciPy-NMeth}
{Virtanen}, P., {Gommers}, R., {Oliphant}, T.~E., {et~al.} 2020, Nature
  Methods, 17, 261, \dodoi{https://doi.org/10.1038/s41592-019-0686-2}

\bibitem[{{Wachlin} {et~al.}(2022){Wachlin}, {Vauclair}, {Vauclair}, \&
  {Althaus}}]{Wachlin_WDs}
{Wachlin}, F.~C., {Vauclair}, G., {Vauclair}, S., \& {Althaus}, L.~G. 2022,
  \aap, 660, A30, \dodoi{10.1051/0004-6361/202142289}

\bibitem[{Warnecke {et~al.}(2023)Warnecke, Korpi-Lagg, Gent, \&
  Rheinhardt}]{warnecke_2023}
Warnecke, J., Korpi-Lagg, M.~J., Gent, F.~A., \& Rheinhardt, M. 2023, Nature
  Astronomy, 7, 662, \dodoi{10.1038/s41550-023-01975-1}

\end{thebibliography}

\end{document}